\documentclass{article}
\usepackage{arxiv}
\usepackage[utf8]{inputenc} 
\usepackage[T1]{fontenc}    
\usepackage{hyperref}       
\usepackage{url}            
\usepackage{booktabs}       
\usepackage{amsfonts}       
\usepackage{nicefrac}       
\usepackage{microtype}      
\usepackage{lipsum}		
\usepackage{graphicx}
\usepackage{natbib}
\usepackage{doi}
\usepackage{float}
\usepackage{graphicx}
\usepackage[ruled,linesnumbered]{algorithm2e}
\usepackage{url}
\usepackage{array}
\usepackage{amsmath}
\usepackage{multirow}
\usepackage{hhline}
\usepackage{float}
\usepackage{tabularray}
\usepackage{titlesec}
\usepackage{comment}
\usepackage{mathtools}
\usepackage{adjustbox}
\usepackage{rotating}

\title{Deep PackGen: A Deep Reinforcement Learning Framework for Adversarial Network Packet Generation}

\date{}

\author{ \href{https://orcid.org/0000-0002-9326-291X}{\includegraphics[scale=0.06]{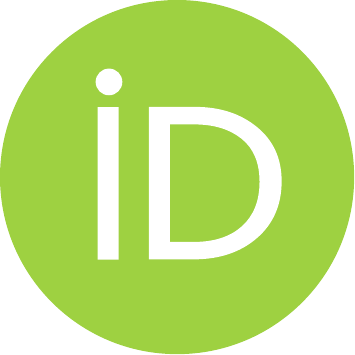}\hspace{1mm}Soumyadeep Hore}\\
	University of South Florida\\
	Tampa, FL 33620 \\
	\texttt{soumyadeep@usf.edu} \\
	\And
    {\hspace{1mm}Jalal Ghadermazi} \\
	University of South Florida\\
	Tampa, FL 33620 \\
	\texttt{jghadermazi@usf.edu} \\
    \And
    {\hspace{1mm}Diwas Paudel} \\
	University of South Florida\\
	Tampa, FL 33620 \\
	\texttt{diwaspaudel@usf.edu} \\
    \And
	{\hspace{1mm}Ankit Shah*} \\
	University of South Florida\\
	Tampa, FL 33620 \\
	\texttt{ankitshah@usf.edu} \\
    \And
    {\hspace{1mm}Tapas K. Das} \\
	University of South Florida\\
	Tampa, FL 33620 \\
	\texttt{das@usf.edu} \\
    \And
	{\hspace{1mm}Nathaniel D. Bastian} \\
	United States Military Academy\\
	West Point, NY 10996 \\
	\texttt{nathaniel.bastian@westpoint.edu} \\
}

\renewcommand{\shorttitle}{Deep PackGen}

\hypersetup{
pdftitle={Deep PackGen: A Deep Reinforcement Learning Framework for Adversarial Network Packet Generation}, 
pdfsubject={},
pdfauthor={Soumyadeep Hore, Jalal Ghadermazi, Diwas Paudel, Ankit Shah, Tapas Das, Nathaniel D. Bastian},
pdfkeywords={Network Intrusion Detection Systems, Adversarial Attack, Deep Reinforcement Learning, DRL Cyber Framework.},
}

\begin{document}
\maketitle

\begin{abstract}
Recent advancements in artificial intelligence (AI) and machine learning (ML) algorithms, coupled with the availability of faster computing infrastructure, have enhanced the security posture of cybersecurity operations centers (defenders) through the development of ML-aided network intrusion detection systems (NIDS). Concurrently, the abilities of adversaries to evade security have also increased with the support of AI/ML models. Therefore, defenders need to proactively prepare for evasion attacks that exploit the detection mechanisms of NIDS. Recent studies have found that the perturbation of flow-based and packet-based features can deceive ML models, but these approaches have limitations. Perturbations made to the flow-based features are difficult to reverse-engineer, while samples generated with perturbations to the packet-based features are not playable.

Our methodological framework, Deep PackGen, employs deep reinforcement learning to generate adversarial packets and aims to overcome the limitations of approaches in the literature. By taking raw malicious network packets as inputs and systematically making perturbations on them, Deep PackGen camouflages them as benign packets while still maintaining their functionality. In our experiments, using publicly available data, Deep PackGen achieved an average adversarial success rate of 66.4\% against various ML models and across different attack types. Our investigation also revealed that more than 45\% of the successful adversarial samples were out-of-distribution packets that evaded the decision boundaries of the classifiers. The knowledge gained from our study on the adversary's ability to make specific evasive perturbations to different types of malicious packets can help defenders enhance the robustness of their NIDS against evolving adversarial attacks.
\end{abstract}


\footnote{This work has been submitted to ACM for possible publication. *Corresponding author}

\section{Introduction}\label{sec:introduction}
A network intrusion detection system (NIDS) is a primary tool for cybersecurity operations centers (CSOCs) to detect cyber-attacks on computer networks. With the availability of high-performance computing resources and advancements in artificial intelligence (AI) and machine learning (ML) algorithms, intrusion detection mechanisms have greatly improved, serving the security needs of organizations. However, adversaries are also continuously advancing their toolchains by using AI/ML-enabled methodologies to camouflage their attacks that can evade these ML-based NIDS. Hence, the CSOCs must improve their security posture by proactively preparing for evasion attacks and making their NIDS robust against evolving adversaries.

\begin{table*}[]
\centering
\caption{Summary of recent literature on adversarial sample generation}
\label{lit_contri}
\resizebox{\textwidth}{!}{
\begin{tabular}{|l|l|l|l|l|l|}
\hline
Author                                                            & Year & Data Set              & Feature     & \begin{tabular}[c]{@{}l@{}}Attacker \\ Knowledge\end{tabular} & Algorithm                                                                                                         \\ \hline
Rigaki et al. \cite{rigaki2017adversarial}       & 2017 & NSL KDD              & Flow-based   & White-box                                                     & \begin{tabular}[c]{@{}l@{}}Fast Gradient Sign \\ Method (FGSM), Jacobian-based \\ Saliency Map Attack (JSMA)\end{tabular}                                                                                                        \\ \hline
Wang et al. \cite{wang2018deep}                  & 2018 & NSL KDD              & Flow-based   & White-box                                                     & \begin{tabular}[c]{@{}l@{}}FGSM, JSMA, Deepfool,\\ Carlini Wagner (CW)\end{tabular}                                                                             \\ \hline
Zhang et al. \cite{zhang2020tiki}                & 2020 & CICIDS-2018          & Flow-based   & White-box                                                     & \begin{tabular}[c]{@{}l@{}}Boundary Attack, Pointwise \\ Attack,  Hopskipjump Attack\end{tabular}                  \\ \hline
Apruzzese et al. \cite{apruzzese2020deep}        & 2020 & CTU, BOTNET          & Flow-based   & Black-box                                                     & Deep Reinforcement Learning                                                                                       \\ \hline
Alhajjar et al. \cite{alhajjar2021adversarial}   & 2020 & \begin{tabular}[c]{@{}l@{}}NSL-KDD,\\ USNW-NB15\end{tabular}   & Flow-based   & Gray-box                                                      & \begin{tabular}[c]{@{}l@{}}Generative Adversarial \\ Network (GAN), Genetic \\ Algorithm (GA), Particle Swarm \\ Optimization\end{tabular}                         \\ \hline
Schneider et al. \cite{schneider2021evaluating}  & 2021 & NSL-KDD              & Flow-based   & White-box                                                     & \begin{tabular}[c]{@{}l@{}}Projected Gradient Descent, GA,\\ Particle Swarm Optimization,\\ GAN\end{tabular} \\ \hline
Chernikova et al. \cite{chernikova2022fence}     & 2022 & CTU 13               & Flow-based   & Black-box                                                     & \begin{tabular}[c]{@{}l@{}}Projected Gradient Descent, CW \end{tabular}                               \\ \hline
Zhang et al. \cite{zhang2022generating}          & 2022 & \begin{tabular}[c]{@{}l@{}}NSL-KDD,\\ UNSW-NB15\end{tabular}   & Flow-based   & Gray-box                                                      & GAN                                                                                                               \\ \hline
Sheatslet et al. \cite{sheatsley2022adversarial} & 2022 & \begin{tabular}[c]{@{}l@{}}NSL-KDD,\\ UNSW-NB15\end{tabular}   & Flow-based   & White-box                                                     & \begin{tabular}[c]{@{}l@{}}Adaptive JSMA, \\ Histogram Sketch Generation\end{tabular}                             \\ \hline
Homoliak et al. \cite{homoliak2018improving}     & 2018 & ASNM-NBPO            & Packet-based & Gray-box                                                      & Tools like NetEM, Metasploit                                                                                   \\ \hline
Hashemi et al. \cite{hashemi2019towards}         & 2019 & CICIDS-2018          & Packet-based & White-box                                                     & Trial and Error                                                                                                   \\ \hline
Kuppa et al. \cite{kuppa2019black}               & 2019 & CICIDS-2018          & Packet-based & Gray-box                                                      & Manifold Approximation                                                                                            \\ \hline
Han et al. \cite{han2021evaluating}              & 2021 & \begin{tabular}[c]{@{}l@{}}Kitsune,\\ CICIDS-2017\end{tabular} & Packet-based & Gray-box                                                      & GAN                                                                                                               \\ \hline
Sharon et al. \cite{sharon2022tantra}            & 2021 & \begin{tabular}[c]{@{}l@{}}Kitsune,\\ CICIDS-2017\end{tabular} & Packet-based & Black-box                                                     & Long Short-Term Memory-based                                                                                                        \\ \hline
\end{tabular}
}
\end{table*}

Evasion attacks on NIDS are mainly conducted by perturbing network flow-based features to deceive ML models. Table~\ref{lit_contri} shows a summary of recent studies that focused on adversarial sample generation to evade NIDS.
However, flow-based attacks are impractical as reverse engineering these perturbations from the flow level into constructing the actual packets is very complex and difficult~\cite{rosenberg2021}. In addition, hidden correlations among different flow-based features further exacerbate the computational difficulty of replaying perturbations in a real network communication~\cite{han2021evaluating}. More importantly, perturbations must be made such that the communication's functionality is maintained. Hence, crafting adversarial attacks at the packet level is necessary to improve the practicality of implementing evasion attacks.

A few studies in recent literature have focused on using packet-based data to construct evasion attacks (see Table~\ref{lit_contri}). These studies have utilized publicly available data sets to obtain the samples for obfuscation and relied on making random perturbations using trial-and-error and other approximation techniques. The generated adversarial samples were then tested against linear, tree-based, and nonlinear ML models for evasion. The limitations of these studies are as follows. The perturbations made to the samples were mainly focused on the time-based features, which a classifier can be made immune to by training it with raw packet information. Some also generate adversarial samples using packet or payload injection and packet damage. However, there exists a correlation among the packet-level features, directly impacting the feature set of the classifier, which is not considered in any of these studies. This phenomenon is also known as the side effect of packet mutation~\cite{pierazzi2020intriguing}. Another limitation of existing packet-based approaches is that they perturb both forward and backward packets (i.e., communication from the host to the destination and then the destination back to the host). Clearly, an adversary can only control the forward packets, those originating from the host and going to the destination (server).

Our proposed methodological framework addresses the above limitations in the following ways. Our methodology uses a learning-based approach, in which an AI agent is trained to make (near-)optimal perturbations to any given malicious packet. The agent learns to make these perturbations in a sequential manner using a deep reinforcement learning (DRL) approach. We identify the forward packets in network communication and only modify them to produce adversarial samples. We evaluate our adversarial samples against classifiers trained using packet-level data. We aim to make minimal and valid perturbations to the original packets that preserve the functionality of the communication. Examples of such perturbations include modifications to the valid portions in the internet protocol (IP) header, transmission control protocol (TCP) header, TCP options, and segment data. Furthermore, we only consider perturbing those features that can be obtained from the raw packet capture (PCAP) files without any preprocessing. This makes it practical to replicate the attack using perturbed packets. We consider the side effects of packet mutation in this study. For example, any change to the IP or TCP header affects the IP and TCP checksum, respectively. A detailed description of the perturbations and their side effects is provided in the numerical experiments section (\#\ref{sec:numexp}). 
We also evaluate whether the learning attained from one environment is transferable to another. We do this to gauge the effectiveness of our methodology in real-world settings where adversaries may not have any knowledge of the ML models and the data used to build the NIDS. We demonstrate the playability of the adversarial packet in a flow using the Wireshark application in the results section (\#\ref{sec:results}) of the paper. In summary, our paper addresses the literature gap for constructing adversarial samples by developing a learning-based methodology with the following characteristics: only the forward packets are perturbed; valid perturbations are considered in order to maintain the functionality of the packets; side effects of perturbations are taken into account; effectiveness of the adversarial samples is tested against unseen classifiers; and demonstrated transferability of the framework to other network environments.

There are several contributions to this research study. The primary contribution is the development of a DRL-enabled methodology capable of generating adversarial network packets for evasion attacks on ML-based NIDS. Our methodological framework, Deep PackGen, takes raw network packets as inputs and generates adversarial samples camouflaged as benign packets. The DRL agent in this framework learns the (near-)optimal policy of perturbations that can be applied to a given malicious network packet, constrained by maintaining its functionality while evading the classifier. To the best of our knowledge, this is the first research study that poses the constrained network packet perturbation problem as a sequential decision-making problem and solves it using a DRL approach. Another novel aspect of this research is creating a packet-based approach to developing classification models for ML-based NIDS. The unidirectional (forward) packets from raw PCAP files are extracted, preprocessed, feature-engineered, and normalized for machine computation. The transformed network packets are then used to train the classifiers. Other contributions highly relevant to the cybersecurity research community include the insights obtained from the experiments and their analyses. Our investigation reveals that our methodology can generate out-of-distribution (OOD) packets that can also evade the decision boundaries of more complex nonlinear classifiers. Furthermore, we also explain why packets of certain attack types can be easily manipulated compared to others. The knowledge gained from this study on the adversary’s ability to make specific
perturbations to different types of malicious packets can be used by the CSOCs to defend against the evolving adversarial attacks.

The rest of the paper is organized as follows. In Section 2, we present related literature pertaining to different types of intrusion detection mechanisms and adversarial attacks on ML models. We also present an overview of DRL approaches used in security and other application domains. Section 3 describes the DRL-enabled Deep PackGen framework for adversarial network packet generation. The data set creation process, packet classification model development and the DRL solution approach are explained in this section. Section 4 discusses the numerical experiments conducted in this study. The performance of our methodological framework on publicly available data sets, the analysis of DRL agent’s policies, and the statistical analysis of the adversarial samples are presented in Section 5. Section 6 presents the insights obtained from this research study, along with the conclusions and future work.

\section{Related Literature}
We first describe the role of NIDS and review its various types, followed by a summary of recent literature on adversarial attacks on ML models and the use of DRL as a solution approach to solving complex problems in various domains.

\subsection{Network Intrusion Detection System}
A NIDS is employed to detect unauthorized activities threatening an information system's confidentiality, integrity, and availability. There are two types of NIDS, signature-based and anomaly-based. A signature-based NIDS matches the signature of an activity with the database comprising signatures of previous malicious activities, while an anomaly-based NIDS models the expected user behavior (benign activity) in a computer and network system to identify any activity outside of this normal behavior (such as a malicious activity). Anomaly-based malicious activity detection can be achieved through ML or statistics-based methods~\cite{lin2015cann}. With advancements in computational technology and availability of faster computing resources, ML is extensively used in anomaly detection, leveraging various techniques such as clustering, deep neural networks, decision trees, and tree-based ensemble algorithms, among many others~\cite{xiao2018iot}, \cite{kshetri2017hacking}. Development of an ML-based NIDS involves fitting a classification model to a training data set containing both malicious and benign data. The trained model is then used to identify malicious activities in the network.

\subsection{Adversarial Attacks on ML Models}
Recent developments in AI/ML algorithms along with reduced computational costs due to mechanisms such as parallel computing have helped organizations better address their security needs~\cite{shone2018deep}. Concurrently, these AI/ML techniques also provide new opportunities for adversaries to launch attacks, circumventing the improvements in the intrusion detection mechanisms. Adversarial ML (AML) is a growing concern in AI research due to the potential security vulnerabilities it can create in ML models. As per the recent technical report published by the National Institute of Standards and Technology ~\cite{oprea2023adversarial}, there are three types of attacks that can be performed on ML models: evasion, poisoning, and privacy attacks. In an evasion attack, the attacker's goal is to generate an adversarial sample by making small perturbations to the original input such that it is misclassified as any other arbitrary class sample. Poisoning attacks are targeted at corrupting the data, model, or labels used during ML model training, aiming to disrupt the availability or integrity of the system. In privacy attacks, the adversary aims to access aggregate statistical information from user records by reverse engineering private user information and critical infrastructure data. Our study focuses on evasion attacks against ML-based models, which pose a big threat to security in an organization.

Adversarial evasion attacks were first observed in computer vision when Szegedy et al. showed that minor changes to an image could cause a deep neural network to misclassify it~\cite{szegedy2013intriguing}. Further research has focused on constructing evasive attacks on various systems in other domains, including attacks on ML-based NIDS in cybersecurity. Evasion attacks on NIDS can be categorized based on the adversary’s knowledge of the classifier~\cite{oprea2023}. \emph{Black-box} attacks consider zero knowledge of the classifier, its hyper-parameters, and the features used during training. \emph{Gray-box} attacks also consider zero knowledge of the classifier but assume some knowledge of the preprocessing functions leading to the training feature set. \emph{White-box} attacks assume complete knowledge of the classifier and its training feature set. The adversarial capability can be defined based on access to the training data. The capability of the attacker can be categorized as \emph{none} if the attacker does not have access to the training data, or \emph{passive} if the attacker has access to both benign and malicious traffic~\cite{he2023adversarial}. In computer vision literature, it is assumed that the adversary cannot access the training data. However, in the case of NIDS, it is reasonable to assume the availability of training data. For instance, an adversary can collect data by silently sniffing traffic in a computer network with the help of software and hardware sniffers. Table~\ref{lit_contri} provides a summary of recent literature studies on evasion attacks by adversarial sample generation against ML-based NIDS. This table shows the data sets used, the types of features exploited, the knowledge of the attacker, and the adversarial methodology utilized in the respective studies.

\subsection{DRL Approaches in Security and Other Fields}
Reinforcement learning (RL) has a high similarity to human learning. In RL, an agent learns optimal actions through experience, by exploring and exploiting the unknown environment. For solving complex real-world problems to near-optimality, a nonlinear function approximator, in the form of a deep-learning model, is integrated within the RL framework. The deep RL or DRL methodology is one of the most successful approaches to solving sequential decision-making problems~\cite{lyu2019knowledge}. 
There are both model-based and model-free DRL approaches. The latter can be further divided into value-based and policy-based algorithms~\cite{sutton2018}. One of the popular value-based approaches is the double Q-Learning (DDQN)~\cite{van2016deep}, which is an enhancement over Mnih's deep Q-learning approach (DQN)~\cite{mnih2015human}. Recent applications of DRL as a solution methodology in various domains include vulnerability management in CSOCs~\cite{HORE2023119734}, optimization of EV sharing systems~\cite{bogyrbayeva2021reinforcement}, training robots to automate complex tasks~\cite{kirtas2020deepbots}, and developing precision advertising strategies~\cite{liang2020precision}, among others.

\section{ADVERSARIAL NETWORK PACKET GENERATION FRAMEWORK: DEEP PACKGEN}

\begin{figure*}
\centering
\includegraphics[width=0.95\textwidth]{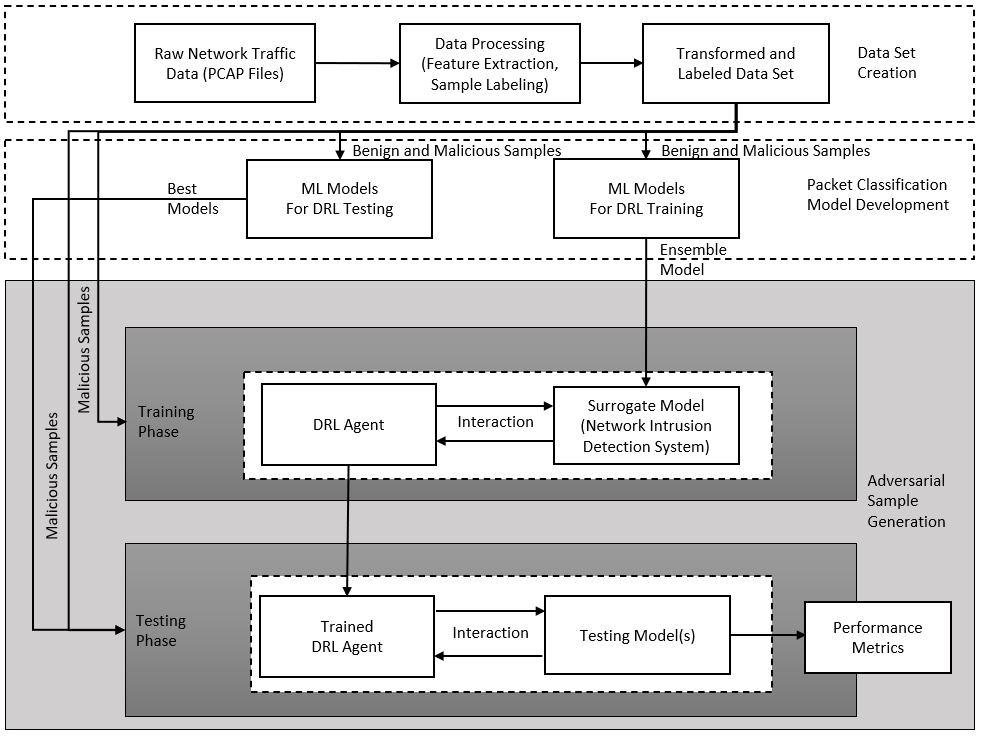}
\caption{Deep PackGen framework for adversarial network packet generation}
\label{framework}
\end{figure*}

The objective of our study is to develop a framework for generating adversarial network packets that can bypass ML-based NIDS while maintaining functionality for communication. Our proposed framework, named Deep PackGen, illustrated in Figure~\ref{framework}, comprises of three main components: data set creation, packet classification model development, and adversarial network packet generation. We begin by describing the process of creating and labeling network traffic data, followed by training and evaluating packet classification models. Finally, we present a DRL model trained to generate adversarial network packets by interacting with several packet classification models.

\subsection{Data Set Creation} \label{datasetcreation}
While much research has been conducted on developing different ML models for network traffic classification, most of it relies on the NIDS (such as Zeek, Snort, and Security Onion) or the NetFlow tools (such as Wireshark and CICFlowmeter) for compiling network packet information to obtain features for training the models. These methods have several limitations as follows: (i) NIDS and NetFlow tools generate features based on predefined rules or signatures, which makes it difficult to reproduce and compare results across different studies; (ii) these approaches often do not incorporate raw payload information, which can make it hard to detect attacks that are embedded in packet payloads; (iii) flow-based features are extracted by analyzing network traffic over a period of time, which makes it difficult to detect anomalies in real-time; and (iv) rule-based and signature-based feature extraction approaches can fail when encountering novel attacks without signatures. To address these limitations, recent research studies have focused on using raw packet data to train ML-based NIDS \cite{lotfollahi2020deep,de2021machine,bierbrauer2023transfer,cheng2021packet}. These studies use bidirectional data to train their ML models. However, an adversary can only control the network packets being sent from one direction (i.e., from the source). Hence, in this study, we create a data set comprising raw packet data with a unidirectional flow originating from the source. The packet data from the unidirectional flows is used to train the ML models and to generate adversarial samples (network packets). We start this process by accessing the PCAP files of publicly available network intrusion data sets, such as the ones by the Canadian Institute of Cybersecurity (CIC) (CICIDS-2017~\cite{sharafaldin2019detailed} and CICIDS-2018~\cite{sharafaldin2018toward}). This process can also be applied to private data sets available at the CSOCs.

\begin{figure}
\centering
\includegraphics[width=0.75\textwidth]{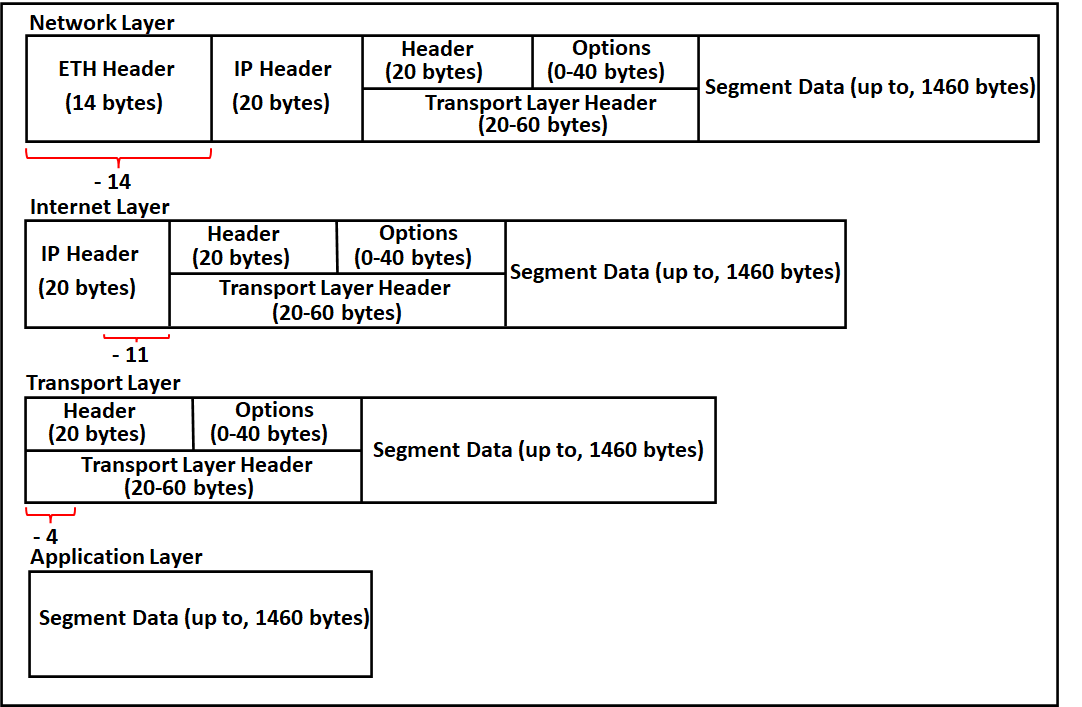}
\caption{TCP/IP model with byte information for each layer}
\label{tcp-ip}
\end{figure}

Our data set creation process uses a tool that extracts raw packet data from the PCAP files, selects unidirectional packets, pre-processes, and transforms the raw data into normalized numeric feature values. The complete procedure for processing and labeling the data is provided in our paper for ease of use. The steps of the procedure are as follows.
\begin{enumerate}
\item Parse the different header and segment information from the PCAP file with the help of parsers such as python-based \textit{Scapy} \cite{biondi2010scapy} and \textit{dpkt} \cite{dpkt}.
\item Label the packet data using the 6-tuple information (source IP address, destination IP address, source port, destination port, protocol, and epoch time) to identify benign and malicious packets. For the malicious packets, each sample is assigned the respective attack class label identified in the historical data set (for instance, from the data described on the CIC website for the CICIDS-2017 and CICIDS-2018 data sets).
\item Extract unidirectional packets using the source IP addresses, i.e., only extracting packets being sent from the identified source, while not including the response from the destination. We will refer to these unidirectional packets as \textit{forward packets}, hereafter.
\item Remove the header information that may increase the bias in training the ML-based classifier. This includes removing the ETH header information, source, and destination IP address information from the IP header, and source and destination port information from the TCP header of each packet data. Figure~\ref{tcp-ip} shows a TCP/IPv4 model with red underlines depicting the location of this information and the number in red fonts representing the number of bytes that will be removed from each layer.
\item Set the feature-length to $N$ for all the packets. Each of the remaining bytes, after the removal of information in the previous step, will be a feature in this data set. The number of bytes will vary based on the type of packet. Hence, to maintain the standard structure of the data, zero-padding is applied to the feature space~\cite{theile2020uav}.
\item Convert the raw information in each packet from hexadecimal to decimal numbers between 0-255 and normalize them to be between 0-1 for efficient machine computation.
\end{enumerate}

\subsection{Packet Classification Model Development}
An adversary may not have complete knowledge of the defender's model. Hence, an adversary will need a substitute for the defender's ML-based NIDS to generate and evaluate the adversarial samples. We propose an ensemble model as a surrogate for the defender's model for training the adversarial agent. An ensemble model consists of multiple estimators (ML models), making the classifier robust in identifying malicious packets. The data set created using the first component of this framework is split into training and testing data sets. Various linear, tree-based, and non-linear ML models are then developed using the training data set and they are evaluated using the testing data set. The selection process of the estimators for this ensemble is described as follows. Given a large set of estimators $E$, in which each estimator is represented as $F_m(.)$, where $m \in E$, then  the best set of estimators, $M$, is selected based on their performance metric values. $F_{m}^{*}(.)$ represents the estimator with optimal parameters for which its loss function value, $L(.)$, is minimum, i.e.,
\begin{equation}
    \theta^{*} = \arg min_{\theta} L(\theta, x, y)
\end{equation}
where, $\theta^{}$ represents the model parameters, $x \in X$ is the training data, $y \in Y$ is the target value, and $L(.)$ measures how far the predictions are from the target value. Finally, the top $|M|$ number of classifiers are selected in the ensemble, representing the defender's ML-based NIDS, shown as follows
\begin{equation}
    F_{m}^{*}(.) ~~\forall ~~ m \in M
\end{equation}

\subsection{Adversarial Sample Generation}
We first provide the general definition of the adversarial sample generation problem, followed by the problem formulation and the DRL-based solution approach.

\subsubsection{Problem Definition}
Our aim is to develop a methodology to generate malicious network packets that can fool the defender's ML-based NIDS. To achieve this, an original malicious packet is perturbed to camouflage it as benign traffic. Unlike the problem of applying unconstrained perturbations to an image to fool a computer vision-based classifier~\cite{wang2020deceiving}, in this problem, the perturbations are constrained by the requirement to maintain the packet's maliciousness and functionality.

An original malicious packet, $x_{original}$, is modified by applying perturbation(s), $\delta$, using a perturbation function, $P(.)$. These perturbations must belong to a set of all valid perturbations, $\Delta$, that do not impede the capability of the packet.  A perturbed sample, $x_{p}$, can be defined as

\begin{equation}
\label{seq}
    x_{p} = P(x_{original},\delta)
\end{equation}

\begin{equation}
    \delta \in \Delta
\end{equation}
Note that many perturbed samples can be obtained by applying different $\delta$ from this set of valid perturbations, resulting in a large set of perturbed samples, $X_p$. However, a successful adversarial sample, $x_{p}^{benign}$, is the malicious and functional network packet in $X_{p}$ that is able to bypass the defender's model by getting misclassified as benign. This can be formally defined as

\begin{equation}
\label{maineqn}
    x_{p}^{benign} = \arg max_{x_{p} \in X_{p}} L(\theta^{*}, x_{p}, y)
\end{equation}

\subsubsection{Problem Formulation}
Generating an adversarial sample by making perturbations to a network packet can be posed as a sequential decision-making problem. An adversary starts with an original malicious network packet and makes sequential perturbations, as indicated in Equation~\ref{seq}. At each iteration, the packet is modified, and this perturbed sample is passed through the packet classification model to check if it successfully evades its classification decision boundary. The iterative process continues until either a successful adversarial sample is attained (satisfying Equation~\ref{maineqn}) or the maximum number of iterations is reached. The objective is to learn the (near-)optimal set of perturbations, given an original malicious network packet, to generate an adversarial sample. This sequential decision-making problem can be formulated as a Markov decision process (MDP). The key elements of the MDP formulation are as follows.

\begin{itemize}
    \item \textbf{State, $s_t$}, is a representation of the information available at time $t$. The state space consists of the normalized byte values of the network packet obtained from the data set created in the first component of this framework and the classification label ($0$ for benign and $1$ for malicious) given by the defender's model. Each packet contains $N$ number of features, which makes the state space $N+1$ dimensional.
    \item \textbf{Action, $a_t$}, represents the perturbation(s), $\delta \in \Delta$, applied to the network packet at time $t$. The number of action choices is limited to $|\Delta|$ and the choices are discrete.
    \item \textbf{Reward, $r_t$}, is the measure of effectiveness of taking action $a_t$ in state $s_t$. The reward signal helps the adversary in quantifying the effect of the action taken in a particular state. We engineer a novel reward function to guide the adversary towards learning an optimal policy of making perturbations, given the original malicious network packet. The reward function is defined as follows:
\begin{equation}
\label{reward}
  r_t(s_t, a_t) = 
  \begin{dcases*} 
  r^{-} & if $y_{m}$ $\neq$ benign $\forall m \in M$ \\
  k * r^{+} & otherwise
  \end{dcases*} 
\end{equation}

where, $k$ is the number of classifiers in the ensemble that were successfully evaded by the perturbed network sample. This function generates both positive and negative rewards. A positive reward is obtained when the perturbed sample evades one or more classifiers in the ensemble model. The reward value is directly proportional to the number of classifiers it is able to fool by getting misclassified as a benign sample. A small negative reward ($r^{-}$) is incurred each time the perturbed sample fails to evade any of the classifiers in the ensemble model.
\end{itemize}

\subsubsection{DRL-based Solution Approach}

The network packet perturbation problem has a large state and action space. To overcome the issue of calculating and storing the action-value (Q value) for all state-action pairs using a conventional RL approach, we use a deep neural network architecture to estimate these values. An adversary, in the form of a DRL agent, is trained using the malicious samples from the data set created in the first component of the framework. It is to be noted that the DRL agent has no visibility of the ML model's architecture, parameter values, or loss function during the training and testing phases. Figure~\ref{framework} shows the training and testing phases of the DRL agent, which are explained next.

\begin{figure}
\centering
\includegraphics[width=0.75\textwidth]{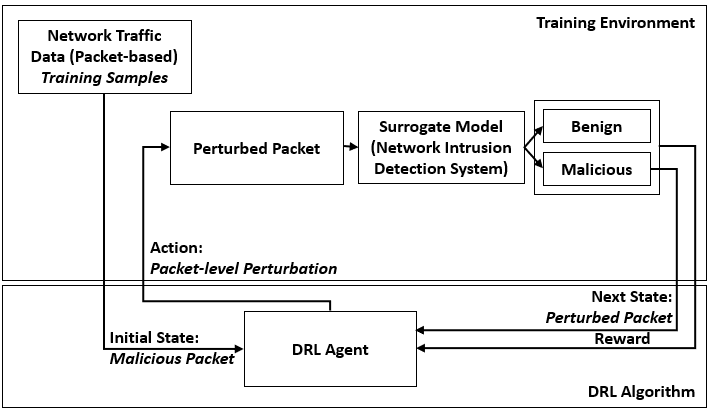}
\caption{Training phase of DRL agent}
\label{drltrain}
\end{figure}

\paragraph{DRL Training Phase}
Figure~\ref{drltrain} shows the training phase of the DRL agent, which comprises interactions between the DRL agent and the training environment. The training environment for the DRL agent is designed with the surrogate for the real-world ML-based NIDS. The environment contains the transformed and labeled network packet data of various attack types, and the pre-trained classifiers for the surrogate ensemble model created in the first and second components of this framework, respectively. For each attack type, the DRL agent obtains a randomly picked data sample and prescribes the perturbation actions (details are explained in the next paragraph). The environment allows for the implementation of these actions, resulting in a one-step transition of the system state, generating a perturbed sample. Rewards are calculated based on the perturbed packet's ability to evade an ensemble of classifiers. This process continues until the stopping condition is reached, which is either the adversarial sample is successful in being misclassified as benign traffic or the maximum number of time-steps is reached. The steps for simulating the environment are outlined in Algorithm 1.

\begin{algorithm}[!t]
\SetAlgoLined
\begin{scriptsize}
\KwIn{Transformed and labeled network packet data for all attack types, pre-trained $|M|$ number of ML classifiers (surrogate ensemble model) for each attack type, total number of time-steps $T$, and total number of episodes $I$.}
\KwOut{System state at time $t+1$, $s_{t+1}$, and reward at time t, $r_t$.}
/* Initiate the training environment simulation */
\caption{Algorithm for simulating the environment}
\For{each attack type in the data set}{
\For{number of episodes, $i<= I$}{
Randomly select one sample from the input data set

\For{number of time-steps, $t <= T$ }{
Obtain action, $a_t$ (from Algorithm 2)

Implement $a_t$ for $s_{t}$ and obtain $s_{t+1}$

Obtain classification, $y_{m} = F_{m}^*(s_{t+1})$ $\forall m \in M$

\If{$y_{m}$ $\neq$ benign $(\forall m \in M)$}{
Obtain a negative reward, $r_t^-$ (Equation~\ref{reward})

Increment $t$ by 1

}\Else{
Obtain a positive reward, $k$ * $r_t^+$ (Equation~\ref{reward})

Update state, $s_{t+1}$ with the updated label value

\If{$y_{m}$ $=$ benign $(\forall m \in M)$}{
Set $s_{t+1}$ as the terminal state

Terminate episode
}
}
}
}
}

\Return {$s_{t+1}$ and $r_t$}
\end{scriptsize}
\label{alg1}
\end{algorithm}

A DRL agent interacts with the training environment and learns to generate adversarial packets by following a set of rules called a policy. The agent's decisions are based on a sequence of states, actions, and rewards, which are determined by the training environment. The agent is rewarded based on the ability of the generated packets to evade the surrogate model. We use DRL with double Q-Learning (DDQN)~\cite{van2016deep} to train the agent. DDQN is a single architecture deep Q-network that is suitable for problems with a discrete action space. The notable difference between DDQN and traditional deep Q-learning is that DDQN decouples the action selection and evaluation processes by using an additional network. This helps to reduce the overestimation error that occurs in traditional Q-learning. The target value calculation is as follows:
\begin{equation}
\label{eq7}
    R_t = r_t + \gamma Q(s_{t+1}, \arg max_a Q(s_{t+1}, a ; \theta_t) ; \theta_t^\prime)
\end{equation}
The policy network weights $(\theta_t)$ are used to select the action, while the target network weights ($\theta_t^\prime$) are used to evaluate it. Algorithm 2 shows the steps in training the DRL algorithm. The algorithm interacts with the environment (Algorithm 1) to learn the near-optimal policy for perturbing different types of packets. The learning process continues until a pre-determined maximum number of episodes is reached. Multiple DRL agents are trained, one for each attack class in the data set.

\begin{algorithm}[!t]
\SetAlgoLined
\begin{scriptsize}
\KwIn{Simulation environment,  total number of time-steps $T$, and the total number of episodes $I$.}
\KwOut{Trained DRL agents for each attack type (final policy network weights).}
/* Initialize policy network $Q_{\theta}$, target network $Q_{\theta^{'}}$, experience replay buffer B. */
\caption{Algorithm for training the DRL agent}
\For{each attack type in the data set}{
\For{number of episodes, $i<= I$}{
Obtain state $s_{t}$ of the randomly selected sample from Algorithm 1

$\epsilon$ $\gets$ setting new epsilon with $\epsilon$-decay

\For{number of time-steps, $t <= T$ }{

Generate action, $a_t$ using $\epsilon$-greedy strategy and pass it to Algorithm 1

Obtain next state, $s_{t+1}$ and reward, $r_{t}$ from Algorithm 1

Store transition in experience replay memory B ($s_{t}, a_{t}, r_{t}, s_{t+1}, done$)

\If{\emph{enough experience} in B}{
Sample a random \emph{minibatch} of b transitions from B

\For{every \emph{minibatch}}{

\If{\emph{done}}{

$R_t = r_t$
}\Else{

$R_t = r_t + \gamma * Q(s_{t+1}, \arg max_{a} Q(s_{t+1}, a ; \theta_t), \theta_{t}^{'})$ (Equation~\ref{eq7})
}
}
Calculate the loss:

$L = 1/|b| \sum_{s_{t} \in b} (Q(s_t, a_t, \theta) - y_t)^2$

Update Q using gradient descent by minimizing L

Copy weights from $Q_{\theta}$ to $Q_{\theta^{'}}$ at every C steps
}
}
}
}
\Return {Trained DRL agents (with final policy network weights $\theta$) for each attack type}
\end{scriptsize}
\label{alg2}
\end{algorithm}

\paragraph{DRL Testing Phase}

\begin{figure}
\centering
\includegraphics[width=0.75\textwidth]{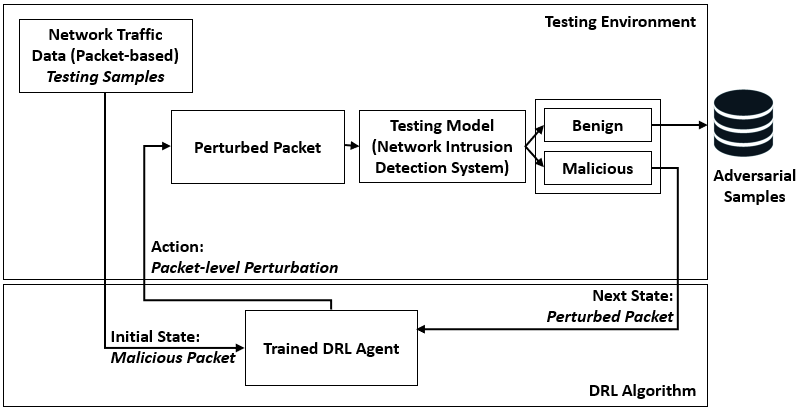}
\caption{Testing phase of DRL agent}
\label{drltest}
\end{figure}

The trained DRL agents are tested against different ML models in the testing phase of the adversarial sample generation component of the framework, as depicted in Figure~\ref{framework}. The neural network architecture of the agent uses the learned weights ($\theta$) obtained at the conclusion of the training phase (see Algorithm 2). The trained DRL agent operates without any reward signal during this phase and its actions are based on its learned policy. Figure~\ref{drltest} illustrates the testing phase of the DRL agent, during which adversarial samples are generated from testing samples that were not seen during the training phase. This allows for an assessment of the agent's performance on various classifiers, including those that were not used in the DRL training environment. The performance of each agent is measured using the adversarial success rate (ASR) metric, which is defined as follows.
\begin{equation}
\label{ASR}
    ASR = \frac{FN_{p} - FN_{original}}{TP}
\end{equation}
where $FN_p$ denotes the total number of samples that were misclassified after perturbation by the DRL agent, $FN_{original}$ is the total number of samples that were incorrectly classified before perturbation, and $TP$ is the total number of samples that the ML model correctly classified as malicious before perturbation. 
The ASR does not take into account packets that fool the classifier prior to perturbation (i.e., $FN_{original}$), thereby giving an accurate performance measurement for each agent.

\section{NUMERICAL EXPERIMENTS}\label{sec:numexp}
In this section, we outline the numerical experiments performed to evaluate our Deep PackGen methodology. We first discuss the experimental data, followed by the creation of ML-based packet classification models. Finally, we delve into the hyperparameters employed during the training and testing of the DRL agent. The goal is to train multiple DRL agents, each specifically designed to generate packets for a unique attack type, by interacting with a surrogate model specialized in identifying that type of attack. To achieve this, we generated several data sets for training and testing of the DRL agents.

\subsection{Data Description}
We conducted numerical experiments using raw PCAP files from two popular network intrusion detection data sets: CICIDS-2017~\cite{sharafaldin2019detailed} and CICIDS-2018~\cite{sharafaldin2018toward}. These data sets contain both benign and attack communications, and provide pragmatic representation of modern network traffic compared to older data sets like NSL-KDD and KDD-CUP \cite{hindy2020taxonomy}. Additionally, the availability of raw PCAP files for the CICIDS data sets reduces dependency on extracted flow level features~\cite{rosenberg2021adversarial}. CICIDS-2017 consists of PCAP files for five consecutive days (Monday to Friday), each with different attack types and sizes, as shown in Table~\ref{tab1}. We processed these files to generate the data set, as explained in Section 3. Table~\ref{tab2} displays the number of forward packets extracted for each attack and its subtypes. Since the Heartbleed and Botnet attack types have too few instances to train ML models, we excluded them from our experiments. As discussed earlier, we only considered forward packets in our data set as an adversary is in control (generation and manipulation) of packets that originate from its source. We extracted payload bytes from each packet and represent each byte as a feature in this data set. We converted hexadecimal numbers to decimal numbers and normalized each feature value to a range of 0-1, where the minimum and maximum feature values were 0 and 255, respectively. In total, there were 1525 features.

We utilized the CICIDS-2017 data set for training and testing our framework as follows. We divided the data into different attack types, including samples from the benign category. For each attack type, we split the data into three parts: 60\% of data for training the DRL agent and building the surrogate model for the training phase, 30\% of data for building other packet classification models for testing the trained DRL agent, and 10\% of data for generating adversarial samples and performing evaluation in the testing phase. In the rest of the paper, we will refer to them as \emph{training}, \emph{ML model testing 1}, and \emph{DRL agent testing 1} data sets, respectively, for each attack type.

\begin{figure}
\centering
\includegraphics[width=0.75\textwidth]{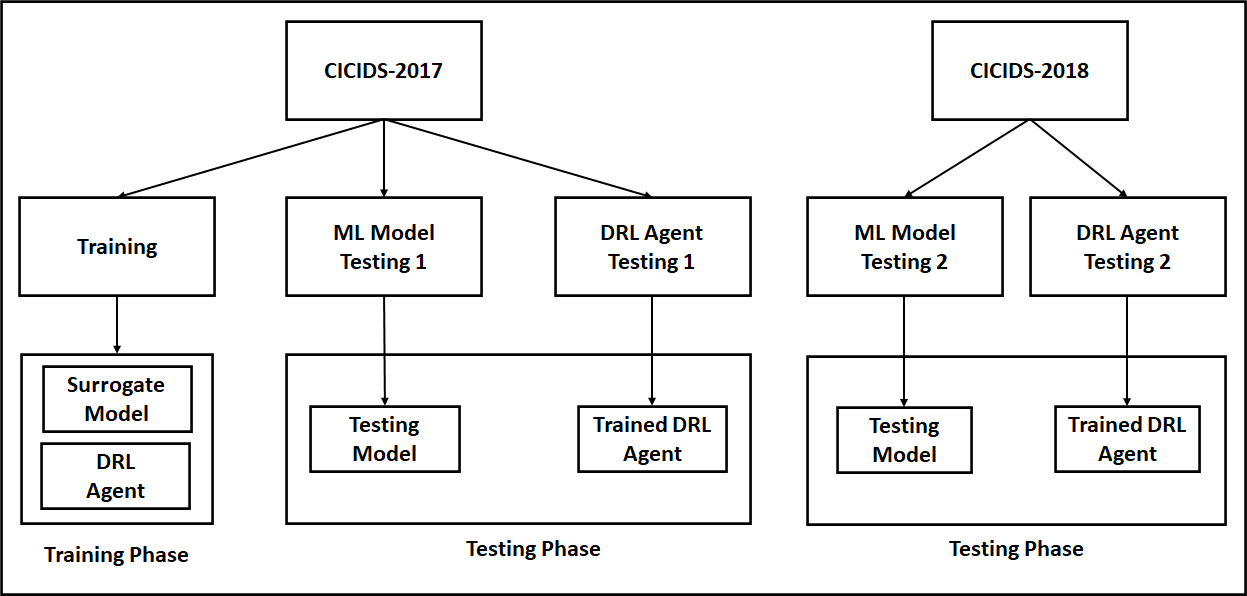}
\caption{Schematic of the data splitting strategy}
\label{data-split}
\end{figure}

Further, to evaluate the performance of the trained agents on different network traffic data, we utilized the CICIDS-2018 data set. Table~\ref{tab3} shows the different attack types and sizes of the PCAP files in the CICIDS-2018 data set. We extracted packets for the various attack types, including DoS, Web Attack, Infiltration, Port Scan, and DDoS from these files. Table~\ref{tab2} displays the number of forward and backward packets for them. We split the data for each attack type into two parts: 70\% of data was allocated for building the ML models for the detection of the respective attack type in the testing phase and the remaining 30\% of data was utilized to measure the adversarial success rate of the respective trained DRL agent. We refer to these two parts as \emph{ML model testing 2} and \emph{DRL agent testing 2} data sets, respectively. Figure~\ref{data-split} shows a schematic of the above-mentioned data splitting strategy from the two CICIDS data sources and their respective uses in the training and testing phases of the framework. Note that the DRL agents were not trained with the CICIDS-2018 data samples.

\begin{table}
\centering
\caption{CICIDS-2017 PCAP file details}
\label{tab1}
\begin{tabular}{|c|l|c|} 
\hline
Day/Date                                                         & \begin{tabular}[c]{@{}l@{}}File \\Size\end{tabular} & Activity                                                                    \\ 
\hline
\begin{tabular}[c]{@{}c@{}}Monday/ \\July 3, 2017\end{tabular}    & 10 GB                                                & Benign                                                                         \\
\hline
\begin{tabular}[c]{@{}c@{}}Tuesday/ \\July 4, 2017\end{tabular}   & 10 GB                                                & \begin{tabular}[c]{@{}c@{}}Brute Force and Benign\end{tabular}               \\
\hline
\begin{tabular}[c]{@{}c@{}}Wednesday/ \\July 5, 2017\end{tabular} & 12 GB                                                & \begin{tabular}[c]{@{}c@{}}DoS and Benign\end{tabular}                       \\ 
\hline
\begin{tabular}[c]{@{}c@{}}Thursday/ \\July 6, 2017\end{tabular}  & 7.7 GB                                               & \begin{tabular}[c]{@{}c@{}}Web Attack, Infiltration, \\and Benign\end{tabular}  \\ 
\hline
\begin{tabular}[c]{@{}c@{}}Friday/ \\July 7, 2017\end{tabular}    & 8.2 GB                                               & \begin{tabular}[c]{@{}c@{}}Botnet, Port Scan, DDoS, \\and Benign\end{tabular}   \\
\hline
\end{tabular}
\end{table}

\begin{table}[]
\centering
\caption{Packet details per attack type in CICIDS-2017 and CICIDS-2018 data sets}
\label{tab2}
\begin{tabular}{|c|c|c|c|}
\hline
\multirow{2}{*}{\begin{tabular}[c]{@{}c@{}}Attack\\ Type\end{tabular}} & \multirow{2}{*}{Subtype}                                & \begin{tabular}[c]{@{}c@{}}Total Packets\\ CICIDS-2017\end{tabular} & \begin{tabular}[c]{@{}c@{}}Total Packets\\ CICIDS-2018\end{tabular} \\ \cline{3-4} 
                                                                       &                                                         & Fwd                                                                 & Fwd                                                                 \\ \hline
\multirow{4}{*}{DoS}                                                   & GoldenEye                                               & 66795                                                               & 160196                                                              \\ \cline{2-4} 
                                                                       & Hulk                                                    & 1246802                                                             & 1026987                                                             \\ \cline{2-4} 
                                                                       & Slowhttptest                                            & 32635                                                               & 105550                                                              \\ \cline{2-4} 
                                                                       & Slowloris                                               & 37236                                                               & 85030                                                               \\ \hline
DDoS                                                                   & -                                                       & 754735                                                              & 546256                                                              \\ \hline
Heartbleed                                                             & -                                                       & 28412                                                               & -                                                                   \\ \hline
Botnet                                                                 & -                                                       & 5788                                                                & -                                                                   \\ \hline
Infiltration                                                           & -                                                       & 29881                                                               & 238087                                                              \\ \hline
Port Scan                                                              &                                                         & 162360                                                              & -                                                                   \\ \hline
\multirow{3}{*}{Web Attack}                                            & \begin{tabular}[c]{@{}c@{}}Brute \\ Force\end{tabular}  & 19755                                                               & 19875                                                               \\ \cline{2-4} 
                                                                       & XSS                                                     & 6361                                                                & 20797                                                               \\ \cline{2-4} 
                                                                       & \begin{tabular}[c]{@{}c@{}}SQL\\ Injection\end{tabular} & 67                                                                  & 334                                                                 \\ \hline
\end{tabular}
\end{table}

\begin{table}
\centering
\caption{CICIDS-2018 PCAP file details}
\label{tab3}
\begin{tabular}{|c|c|c|} 
\hline
Day/Date                                                         & \begin{tabular}[c]{@{}c@{}}File \\Size\end{tabular} & Activity                                                           \\ 
\hline
\begin{tabular}[c]{@{}c@{}}Wednesday/ \\Feb 14, 2018\end{tabular} & 147 GB                                               & \begin{tabular}[c]{@{}c@{}}SSH and FTP Patator, \\and Benign\end{tabular}          \\ 
\hline
\begin{tabular}[c]{@{}c@{}}Thursday/ \\Feb 15, 2018\end{tabular}  & 57.8 GB                                              & \begin{tabular}[c]{@{}c@{}}DoS Goldeneye, DoS Slowloris,\\and Benign\end{tabular}  \\
\hline
\begin{tabular}[c]{@{}c@{}}Friday/ \\Feb 16, 2018\end{tabular}    & 460 GB                                               & \begin{tabular}[c]{@{}c@{}}DoS Slowhttptest, DoS Hulk, \\and Benign\end{tabular}    \\ 
\hline
\begin{tabular}[c]{@{}c@{}}Wednesday/ \\Feb 21, 2018\end{tabular} & 97.5 GB                                              & \begin{tabular}[c]{@{}c@{}}DDoS and Benign\end{tabular}              \\
\hline
\begin{tabular}[c]{@{}c@{}}Thursday/ \\Feb~ 22, 2018\end{tabular} & 110 GB                                               & \begin{tabular}[c]{@{}c@{}}Web Attack and Benign\end{tabular}        \\ 
\hline
\begin{tabular}[c]{@{}c@{}}Friday/\\Feb 23,2018\end{tabular}      & 65.9 GB                                              & \begin{tabular}[c]{@{}c@{}}Web Attack and Benign\end{tabular}       \\ 
\hline
\begin{tabular}[c]{@{}c@{}}Wednesday/\\Feb 28, 2018\end{tabular}  & 73.1 GB                                              & \begin{tabular}[c]{@{}c@{}}Infiltration and Benign\end{tabular}      \\
\hline
\end{tabular}
\end{table}

\subsection{Packet Classification Model Creation}
We developed different sets of ML models for both training and testing the DRL agents. Data sets were carefully prepared for developing the surrogate (training phase) and the testing (testing phase) models. ML models in the training phase were developed using the \emph{training} data. In contrast, those used for testing the trained DRL agents were developed using either the \emph{ML model testing 1} or \emph{ML model testing 2} data sets (see Figure~\ref{data-split}). A DRL agent was trained to perturb packets for each attack type. To effectively train the agent to deceive the classifier's decision boundary, we used a surrogate model specializing in that respective attack type in the agent's training environment. For example, if the DRL agent was being trained on perturbing the packets of a \emph{Port Scan} attack, then the surrogate model was trained with the forward packets extracted from the network flow data of the same attack.

In the training phase, we selected an ensemble of ML models to act as a surrogate model. We randomly sampled 80\% of the \emph{training} data to train various ML models, including linear, tree-based, and nonlinear classification models. The performances of all these models on the remaining 20\% of the \emph{training} data were comparable across each attack type, with the majority of them having a superior accuracy of around 99\%. We selected one model from each of the three types of classifiers in the ensemble: logistic regression (LR), decision tree (DT), and multi-layer perceptron (MLP). Table~\ref{ensemble-accuracy} shows the accuracy scores of these models that form the ensemble in the training environment.

\begin{table}
\centering
\caption{Accuracy scores for models selected in the ensemble}
\label{ensemble-accuracy}
\begin{tblr}{
  row{odd} = {c},
  row{4} = {c},
  row{6} = {c},
  row{8} = {c},
  cell{1}{2} = {r=2}{},
  cell{1}{3} = {r=2}{},
  cell{1}{4} = {r=2}{},
  vlines,
  hline{1,3-9} = {-}{},
  hline{2} = {1}{},
}
Classifiers  & LR   & DT    & MLP  \\
Attack Type  &      &       &      \\
DoS          & 99\% & 99\%  & 99\% \\
DDoS         & 96\% & 99\% & 99\% \\
Web Attack   & 99\% & 99\% & 99\% \\
Port Scan    & 99\% & 99\% & 99\% \\
Infiltration & 99\% & 99\%  & 99\% 
\end{tblr}
\end{table}

Similarly, ML models were developed for the testing phase. Two sets of models were trained: one using the \emph{ML model testing 1} data set and another using the \emph{ML model testing 2} data set. Note that both these data sets contain previously unseen samples by the DRL agents. In addition, the latter contains samples from a different network than that used to train the agents. We used a similar split of 80\% on these data sets to train the respective sets of models. The hyperparameters used for training the various models, including random forest (RF), deep neural network (DNN), and support vector machine (SVM), among others, are shown in Table~\ref{hp}. The values of some of the hyperparameters were adopted from literature \cite{hore2022towards},\cite{ghadermazi2021adversarial}, and others were experimentally determined. The testing accuracy scores of a sample list of models for each attack type on the remaining 20\% of the respective data sets are shown in Table~\ref{defperf}. All these models performed very well in accurately differentiating samples between the benign and malicious classes.

\begin{table*}
\centering
\caption{Hyperparameter values for ML models}
\label{hp}
\begin{tabular}{|c|c|c|c|c|c|} 
\hline
Classifier            & \multicolumn{5}{c|}{Hyperparameters}                                                                 \\ 

\hline
DT          & criterion                    & max\_depth    & min\_samples\_split & max\_features & ccp\_alpha      \\ 
\hline
                 & gini                       & 1500          & 1                   & 39        & 0.05            \\ 
\hline
RF          & n\_estimators                & max\_features & ccp\_alpha          & criterion     & max\_depth      \\ 
\hline
                 & 200                          & sqrt        & 0.04                & gini        & 100             \\ 
\hline
MLP  & hidden\_layer\_sizes         & activation    & solver              & batch\_size   & learning\_rate  \\ 
\hline
                 & (100,)                       & relu        & adam              & 200        & constant      \\ 
\hline
DNN    & hidden\_layer\_sizes         & activation    & solver              & batch\_size   & learning\_rate  \\ 
\hline
                 & (256,128,32,)                & relu        & adam              & 200        & constant      \\ 
\hline
SVM & C (regularization parameter) & kernel        & degree              & gamma         & -               \\ 
\hline
                 & 1.0                          & rbf         & 3                   & scale       & -               \\
\hline
\end{tabular}
\end{table*}

\begin{table*}
\centering
\caption{Testing accuracy scores of trained ML models for testing phase}
\label{defperf}
\label{tab5}
\begin{tabular}{|c|c|c|c|c|c|c|c|c|c|c|} 
\hline
Data Set                           & \multicolumn{5}{c|}{CICIDS-2017}                                                                                & \multicolumn{5}{c|}{CICIDS-2018}                                                                                 \\ 
\hline
Classifier                       & \multirow{2}{*}{DT} & \multirow{2}{*}{RF} & \multirow{2}{*}{MLP} & \multirow{2}{*}{DNN} & \multirow{2}{*}{SVM} & \multirow{2}{*}{DT}  & \multirow{2}{*}{RF} & \multirow{2}{*}{MLP} & \multirow{2}{*}{DNN} & \multirow{2}{*}{SVM} \\ 
\cline{1-1}
\multicolumn{1}{|l|}{Attack Type} &                     &                     &                      &                      &                      &                     &                      &                      &                      &                      \\ 
\hline
DoS                               & 99\%                & 99\%                & 99\%                 & 99\%                 & 99\%                 & 99\%                 & 99\%                & 99\%                 & 99\%                 & 98\%                 \\ 
\hline
DDoS                              & 99\%                & 99\%               & 99\%                 & 99\%                 & 99\%                 & 99\%                & 99\%                & 99\%                 & 99\%                 & 98\%                 \\ 
\hline
Web Attack                        & 99\%                & 99\%               & 99\%                 & 99\%                 & 99\%                 & 99\%                 & 99\%                & 99\%                 & 99\%                 & 88\%                 \\ 
\hline
Port Scan                         & 99\%                & 99\%               & 99\%                 & 99\%                 & 99\%                 & -                    & -                   & -                    & -                    & -                    \\ 
\hline
Infiltration                      & 99\%                & 99\%                & 99\%                 & 99\%                 & 99\%                 & 99\%                 & 98\%                & 97\%                 & 98\%                 & 90\%                 \\
\hline
\end{tabular}
\end{table*}

\subsection{Adversarial DRL Agent Training}
The state space of the DRL agent consists of the 1525 features extracted from the network packet and its classification label. As discussed in the data set creation component (Section~\ref{datasetcreation}) of the framework, these features are the normalized values of the bytes pertaining to different TCP/IP header and segment information (see Figure~\ref{tcp-ip}). Our focus in this study is to find the (near-)optimal set of perturbations that can be applied to a given malicious network packet to generate a successful adversarial sample, while maintaining the functionality of communication. To show the effectiveness of our methodology, we selected a set of valid perturbations ($\Delta$) based on domain knowledge, literature studies~\cite{nasr2020blind}, \cite{sadeghzadeh2021adversarial}, \cite{yan2019automatically}, \cite{guo2021black}, \cite{apruzzese2020deep}, \cite{huang2020adversarial}, and our discussions with the subject matter experts (security personnel) at a collaborating CSOC. Below is a sample list of perturbations, among others, that were selected as a part of the agent’s action space along with their descriptions and impacts.

\begin{itemize}
\item Modifying the fragmentation bytes from \emph{do not fragment} to \emph{do fragment}. This perturbation can be applied to packets where fragmentation is turned off. The \emph{do fragment} command takes the hexadecimal value $40$. This perturbation directly affects byte numbers 7 and 8, and indirectly affects byte numbers 9, 11, and 12 of the IP header, where byte 9 represents the time to live (TTL) value, and bytes 11 and 12 represent the IP checksum value. IP checksum can be calculated by adding all the elements present in the IP header skipping only the checksum bytes \cite{checksum}. The checksum value changes with the change in the byte value of the fragmentation bytes. The TTL value is also adjusted as a result of fragmentation {\cite{stevens1994tcp}}.
\item Modifying the fragmentation bytes from \emph{do not fragment} to \emph{more fragment}. This perturbation can be applied to packets where fragmentation is turned on or off. The \emph{more fragment} command takes the hexadecimal value $20$. This perturbation directly affects byte numbers 7 and 8, and indirectly affects byte numbers 9, 11, and 12 of the IP header.
\item Increasing or decreasing (+/- 1) the TTL byte value. Any valid perturbation to this byte will result in a final TTL value between 1-255. This perturbation directly affects byte number 9, and indirectly affects bytes 11 and 12 of the IP header.
\item Increasing or decreasing (+/- 1) the window size bytes. Any valid perturbation to these bytes will result in a final window size value between 1-65535. This directly impacts byte numbers 15 and 16 of the TCP header, and indirectly impacts byte 17 and 18 of the TCP header, which represent the TCP checksum, similar to IP checksum~\cite{cerf1974protocol}.
\item Adding, increasing, or decreasing the maximum segment size (MSS) value. This perturbation can only be applied to SYN and SYN-ACK packets. For packets that already have the MSS option, we only increase/decrease the value. The MSS value is limited between 0-65535. The TCP options do not have a specific order, and the MSS options have a length of 2 or 4 bytes. This perturbation also indirectly affects byte numbers 17 and 18 of the TCP header.
\item Adding, increasing, or decreasing the window scale value. This perturbation can only be applied to SYN and SYN-ACK packets. For packets that do not have the window scale by default, we add the window scaling to the SYN or SYN-ACK packets, while for packets that already have the window scale option, we only increase/decrease the value. The window scale value is limited between 0-14. The window scale options have a length of 1 or 2 bytes. This perturbation also indirectly affects byte numbers 17 and 18 of the TCP header.
\item Adding segment information. For this perturbation, we selected the most commonly occurring TCP payload information from the benign traffic in the data set. Each time this action is chosen, a portion of the above TCP payload is sequentially added as dead bytes to the end of the malicious packet's TCP payload information ~\cite{aiken2019investigating}, \cite{apruzzese2020deep}, \cite{shu2020generative}. This perturbation affects byte values in the TCP segment, and indirectly affects byte numbers 17 and 18 of the TCP header.
\end{itemize}

We tried various reward schemes with different values of positive and negative rewards in our reward function (see Equation~\ref{reward}). We obtained the best results, in terms of higher average reward value and a faster convergence with the reward term values as follows. We assigned a value of 200 to $r^+$ and -2  to $r^-$. If the perturbed sample successfully evaded all three classifiers in the ensemble model, then $r_t=600$ was passed on to the DRL agent. If the sample successfully evaded only two (one) of the three classifiers, then the DRL agent received a reward of 400 (200). However, if the sample failed to evade any of the classifiers, then a negative reward of $r_t=-2$ was assigned to the agent at time $t$.

\begin{figure}
\centering
\includegraphics[width=0.75\textwidth]{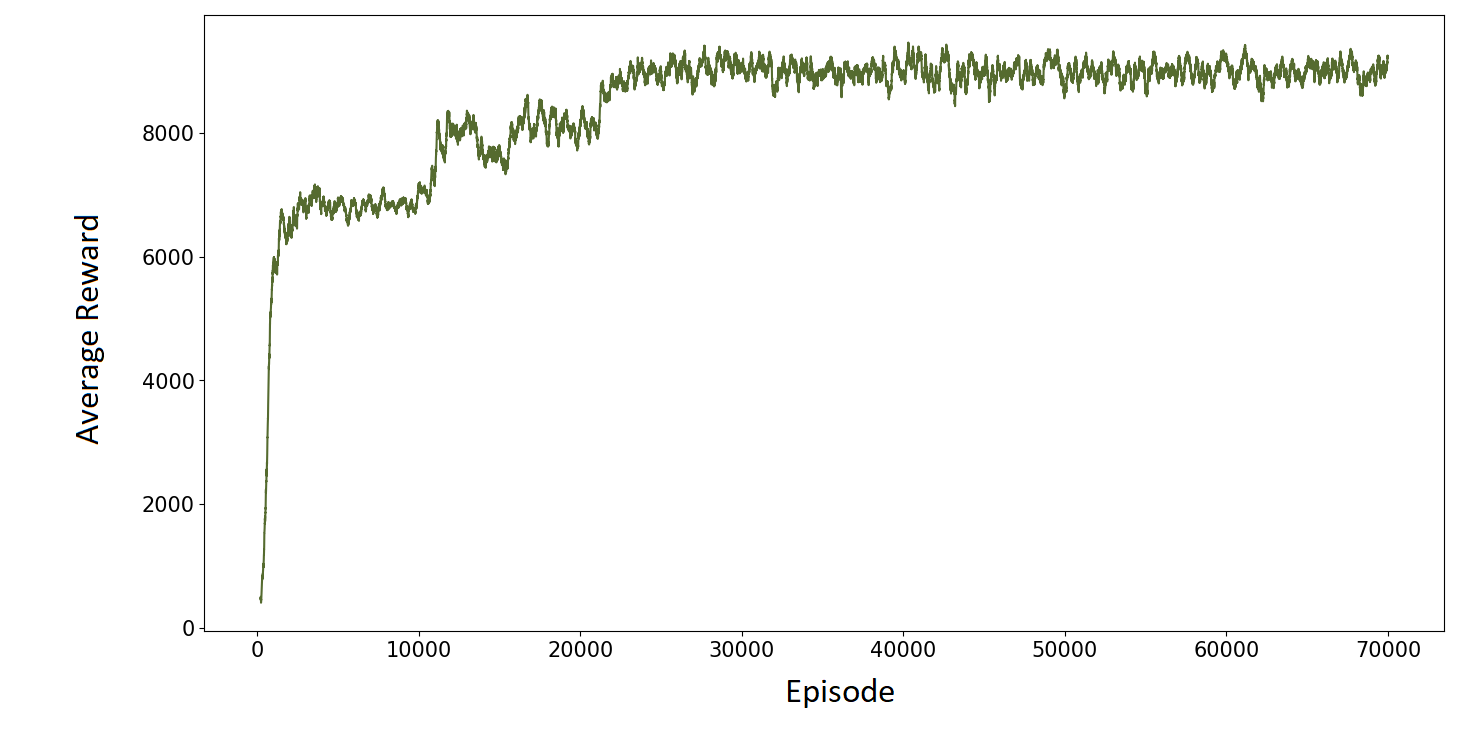}
\caption{A simple moving average reward curve of Port Scan agent during training}
\label{learn}
\end{figure}

We conducted the experiments on a machine with $12^{th}$ Generation Intel Core i9-12950HX processor (30 MB cache, 24 threads, 16 cores) with NVIDIA RTX A5500 graphics card (16GB GDDR6 SDRAM). Table~\ref{tab7} shows the different hyperparameter values used in Algorithm 2 for training the DRL agent. We performed controlled exploration by employing an $\epsilon$-greedy exploration approach with exponentially decaying value of $\epsilon$. The policy network and the target network implemented in the experiments are similar, and the latter is updated with the policy network weights every 10 time-steps. To expedite convergence, Kaiming normal initialization is used instead of random initialization for neural network weights~\cite{he2015delving}. The maximum number of training episodes is chosen based on the improvement curve of the moving average of episodic rewards observed during DRL agent training. Figure~\ref{learn} depicts one such moving average curve during the training of a DRL agent aimed at generating adversarial packets using the \emph{training} data samples for the Port Scan attack type. We noted that the average reward plateaus around 50K episodes across all attack types during the training phase.

\begin{table}
\centering
\caption{Hyperparameter values for Algorithm 2}
\label{tab7}
\begin{tabular}{|l|l|} 
\hline
Hyperparameter            & Value                              \\ 
\hline
No. of Training Episodes  & 50000                               \\ 
\hline
Max. Episode Length        & 30                                  \\ 
\hline
Batch Size                & 256                                 \\ 
\hline
Gamma                     & 0.8                                 \\ 
\hline
Exploration Strategy      & $\epsilon$-greedy  \\ 
\hline
Epsilon Start             & 1                                   \\ 
\hline
Epsilon End               & 0.01                                \\ 
\hline
Epsilon Decay             & 0.00002                             \\ 
\hline
Memory Size               & 100000                              \\ 
\hline
Learning Rate             & 0.001                               \\ 
\hline
Target Update Frequency   & 10                                  \\ 
\hline
Number of Hidden Layers   & 3                                   \\ 
\hline
Hidden Layer Architecture & (256,128,64,)                       \\ 
\hline
Activation                & Relu                                \\ 
\hline
Weight Initialization     & Kaiming Normal                      \\
\hline
\end{tabular}
\end{table}

\section{RESULTS AND ANALYSIS}\label{sec:results}
This section discusses the results of the conducted experiments and their analysis. First, we present the performances of the trained DRL agents against the various testing models. We then present an analysis of the agents' decision-making. Finally, we delve into a deeper statistical analysis of the successful adversarial samples.

\subsection{Performance Evaluation of the Trained DRL Agents}
We evaluate the performance of the trained agents using the \emph{DRL agent testing 1} (CICIDS-2017) data set. We quantify their performance by calculating the rate of adversarial samples that successfully bypass the classification boundary of each testing model by getting misclassified as benign. We use the ASR metric (see Equation~\ref{ASR}) to report each agent's performance. Note that in calculating this performance metric value, the packets that fool the classifier prior to perturbation (i.e., $FN_{original}$) are disregarded (subtracted) to accurately measure the effectiveness of the agent's learned policy.

Table~\ref{tab:ASR-CICIDS-2017} presents the ASR values for the five trained DRL agents (one for each attack type) on five testing models trained using the \emph{ML model testing 1} (CICIDS-2017) data set. The average ASR value obtained across all DRL agents and testing models was 0.664. The DT classifier was found to be the easiest to fool by all agents, with ASR values greater than 0.96 (as shown in the DT column of Table~\ref{tab:ASR-CICIDS-2017}). The DRL agents also performed well against the RF classifier, a tree-based ensemble model, with an average ASR value of 0.694. In particular, the \emph{DDoS} and \emph{Port Scan} agents had similar success rates against both DT and RF classifiers. Some DRL agents, such as \emph{Infiltration} and \emph{Web Attack}, had lower success rates against more complex nonlinear models, such as DNN and SVM classifiers. In general, we observed that simpler models were easier for the DRL agents to evade the decision boundary through adversarial sample generation. The \emph{DDoS} and \emph{Port Scan} agents performed the best against all types of models, while the perturbed packets generated by the \emph{Infiltration} agent had a lower success rate in fooling the nonlinear classifiers.

\begin{figure}
\centering
\includegraphics[width=0.60\textwidth]{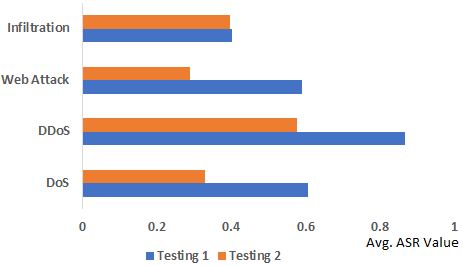}
\caption{Average ASR values of DRL agents in different testing environments}
\label{avgASR}
\end{figure}

Next, we evaluate the transferability of the learned policies of the DRL agents to a different environment (testing 2). To accomplish this, we employ the CICIDS-2018 data set. The five testing models are trained using the \emph{ML model testing 2} data samples. The DRL agents, which were trained using the CICIDS-2017 data, are subjected to malicious samples from the CICIDS-2018 data set (i.e., \emph{DRL agent testing 2} data samples). Note that there are no Port Scan attack samples in this data set, so we present the evaluation of the other four DRL agents in this environment. Table~\ref{tab:ASR-CICIDS-2018} shows the transferability evaluation of these agents using the ASR metric. Overall, the agents successfully perturb malicious samples that evade the classification boundaries of the various testing models in this new environment, with an average ASR value of 0.398. We observed that the DRL agents were more successful in fooling the tree-based models than the nonlinear models. Notably, the \emph{DDoS} agent performed the best, consistent with the findings of the testing 1 environment. Figure~\ref{avgASR} depicts the average ASR values of the DRL agents in both testing 1 and testing 2 environments, highlighting their performance in generating successful adversarial samples.

\begin{table}[]
\centering
\caption{Adversarial success rate (ASR) of DRL agents on CICIDS-2017 data}
\label{tab:ASR-CICIDS-2017}
\begin{tabular}{|c|ccccc|}
\hline
\multirow{3}{*}{Attack Type} & \multicolumn{5}{c|}{Testing 1 Models}                                                                                                                                                                   \\ \cline{2-6} 
                             & \multicolumn{1}{c|}{\multirow{2}{*}{DT}} & \multicolumn{1}{c|}{\multirow{2}{*}{RF}} & \multicolumn{1}{c|}{\multirow{2}{*}{MLP}} & \multicolumn{1}{c|}{\multirow{2}{*}{DNN}} & \multirow{2}{*}{SVM} \\
                             & \multicolumn{1}{c|}{}                    & \multicolumn{1}{c|}{}                    & \multicolumn{1}{c|}{}                     & \multicolumn{1}{c|}{}                     &                      \\ \hline
DoS                          & \multicolumn{1}{c|}{0.969}               & \multicolumn{1}{c|}{0.539}               & \multicolumn{1}{c|}{0.472}                & \multicolumn{1}{c|}{0.786}                & 0.272                \\ \hline
DDoS                         & \multicolumn{1}{c|}{0.965}               & \multicolumn{1}{c|}{0.965}               & \multicolumn{1}{c|}{0.990}                   & \multicolumn{1}{c|}{0.679}                 & 0.742                \\ \hline
Web Attack                   & \multicolumn{1}{c|}{0.995}               & \multicolumn{1}{c|}{0.656}               & \multicolumn{1}{c|}{0.654}                & \multicolumn{1}{c|}{0.330}                   & 0.322                \\ \hline
Port Scan                    & \multicolumn{1}{c|}{0.995}               & \multicolumn{1}{c|}{0.985}                & \multicolumn{1}{c|}{0.979}                 & \multicolumn{1}{c|}{0.314}                & 0.982                \\ \hline
Infiltration                 & \multicolumn{1}{c|}{0.998}               & \multicolumn{1}{c|}{0.324}               & \multicolumn{1}{c|}{0.209}                & \multicolumn{1}{c|}{0.158}                 & 0.323                \\ \hline
\end{tabular}
\end{table}

\begin{table}[]
\centering
\caption{Transferability evaluation (ASR) of DRL agents on CICIDS-2018 data}
\label{tab:ASR-CICIDS-2018}
\begin{tabular}{|c|ccccc|}
\hline
\multirow{3}{*}{Attack Type} & \multicolumn{5}{c|}{Testing 2 Models}                                                                                                                                                                   \\ \cline{2-6} 
                             & \multicolumn{1}{c|}{\multirow{2}{*}{DT}} & \multicolumn{1}{c|}{\multirow{2}{*}{RF}} & \multicolumn{1}{c|}{\multirow{2}{*}{MLP}} & \multicolumn{1}{c|}{\multirow{2}{*}{DNN}} & \multirow{2}{*}{SVM} \\
                             & \multicolumn{1}{c|}{}                    & \multicolumn{1}{c|}{}                    & \multicolumn{1}{c|}{}                     & \multicolumn{1}{c|}{}                     &                      \\ \hline
DoS                          & \multicolumn{1}{c|}{0.506}               & \multicolumn{1}{c|}{0.455}                & \multicolumn{1}{c|}{0.355}                & \multicolumn{1}{c|}{0.225}                & 0.104                \\ \hline
DDoS                         & \multicolumn{1}{c|}{0.657}                & \multicolumn{1}{c|}{0.810}                  & \multicolumn{1}{c|}{0.372}                & \multicolumn{1}{c|}{0.679}                 & 0.372                \\ \hline
Web Attack                   & \multicolumn{1}{c|}{0.312}               & \multicolumn{1}{c|}{0.298}               & \multicolumn{1}{c|}{0.283}                & \multicolumn{1}{c|}{0.293}                & 0.254                \\ \hline
Infiltration                 & \multicolumn{1}{c|}{0.293}               & \multicolumn{1}{c|}{0.354}                & \multicolumn{1}{c|}{0.590}                & \multicolumn{1}{c|}{0.410}                & 0.340                \\ \hline
\end{tabular}
\end{table}

\subsection{Performance Analysis of the Agents}
\begin{figure*}
\centering
\includegraphics[width=0.95\textwidth]{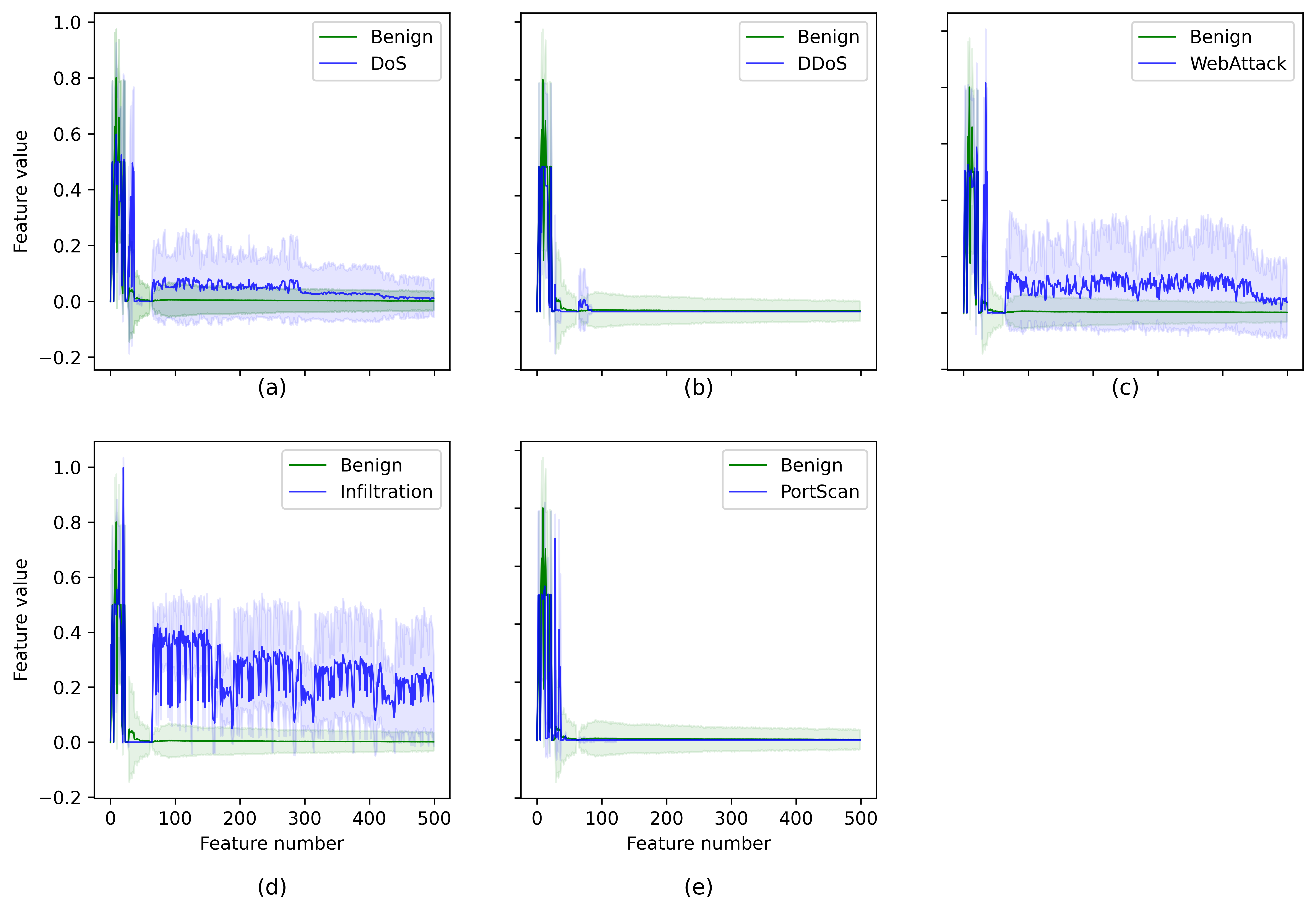}
\caption{Mean and +/-1 standard deviation values for the first 500 features of benign and malicious packets: (a) DoS, (b) DDoS, (c) Web Attack, (d) Infiltration, and (e) Port Scan}
\label{benignattack}
\end{figure*}

\begin{figure*}
\centering
\includegraphics[width=0.99\textwidth, height=0.35\textheight]{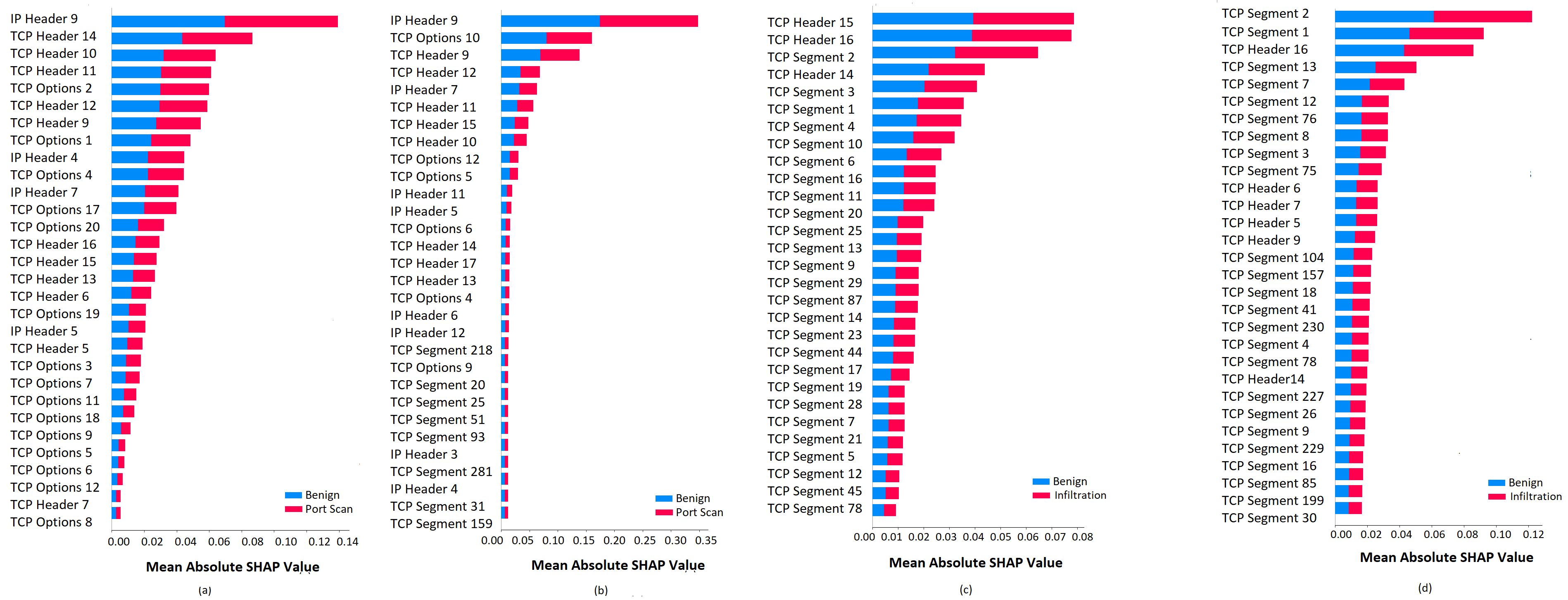}
\caption{SHAP values for (a) RF (Port Scan), (b) MLP (Port Scan), (c) RF (Infiltration), and (d) MLP (Infiltration)}
\label{SHAP}
\end{figure*}

We now assess the effectiveness of the DRL agents by examining two key aspects. First, we analyze why packets of certain attack types were easier to perturb than others. Second, we examine the (near-)optimal actions learned by the agents and their relevance to the decision boundary of the classifiers. To accomplish this, we implemented the following two steps. (i) We calculated the mean and standard deviation values of the first 500 normalized features from the network packets of each attack type in the CICIDS-2017 data set. These values were then compared with those obtained from the benign packets to determine similarity (or dissimilarity) in feature values. This information is visualized in Figure~\ref{benignattack}, which displays the plots for the five attack types. (ii) We used SHapley Additive exPlanations (SHAP)~\cite{lundberg2017unified} to identify important features that determine the classification boundary for accurately detecting each attack type in the testing models. We analyze the agent's performance and action choices in relation to these key features and the ASR values reported in Table~\ref{tab:ASR-CICIDS-2017}. Figure~\ref{SHAP} (a--d) shows the mean absolute SHAP values of the top features found in two classifiers, RF and MLP, developed for detecting two of the attack types, Port Scan and Infiltration.

From Figure~\ref{benignattack} and Table~\ref{tab:ASR-CICIDS-2017}, we can infer that attack types with packet feature values similar to those of the benign class are more susceptible to successful perturbation, allowing evasion of the classifiers. Specifically, plots (b) and (e) in Figure~\ref{benignattack} show that the malicious packets from DDoS and Port Scan attacks, respectively, have similar feature values as the benign packets, resulting in higher ASR values for the DRL agents trained to perturb them, as seen in Table~\ref{tab:ASR-CICIDS-2017}. Next, we look at the agent's actions and important features of the classifier, and analyze how that impacted the ASR.

\begin{figure*}
\centering
\includegraphics[width=0.95\textwidth, height=0.5\textheight]{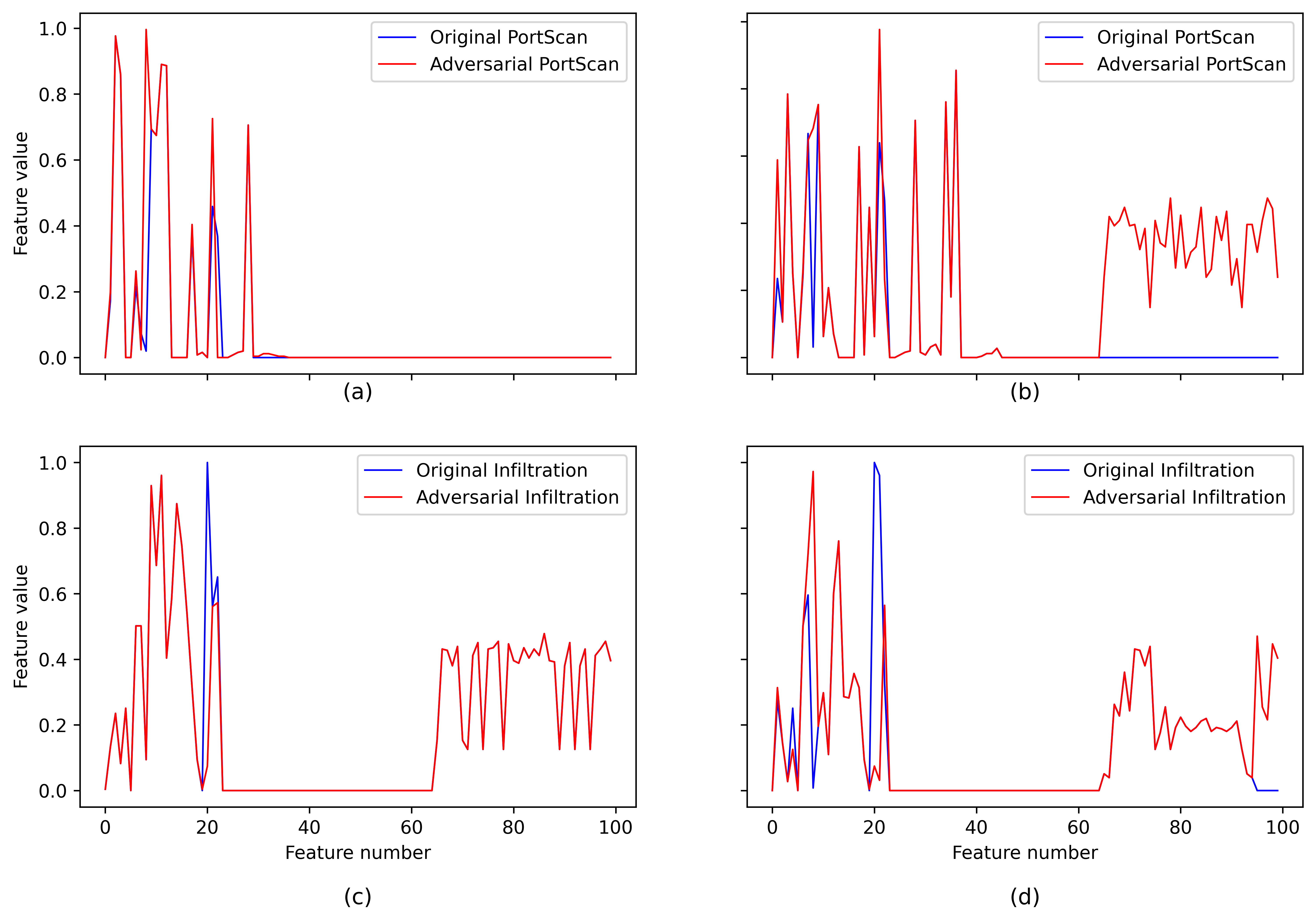}
\caption{Comparison of first 100 features of an original malicious packet and related adversarial sample: (a--b) Port Scan and (c--d) Infiltration}
\label{origpert}
\end{figure*}

\begin{figure*}
\centering
\includegraphics[width=0.95\textwidth]{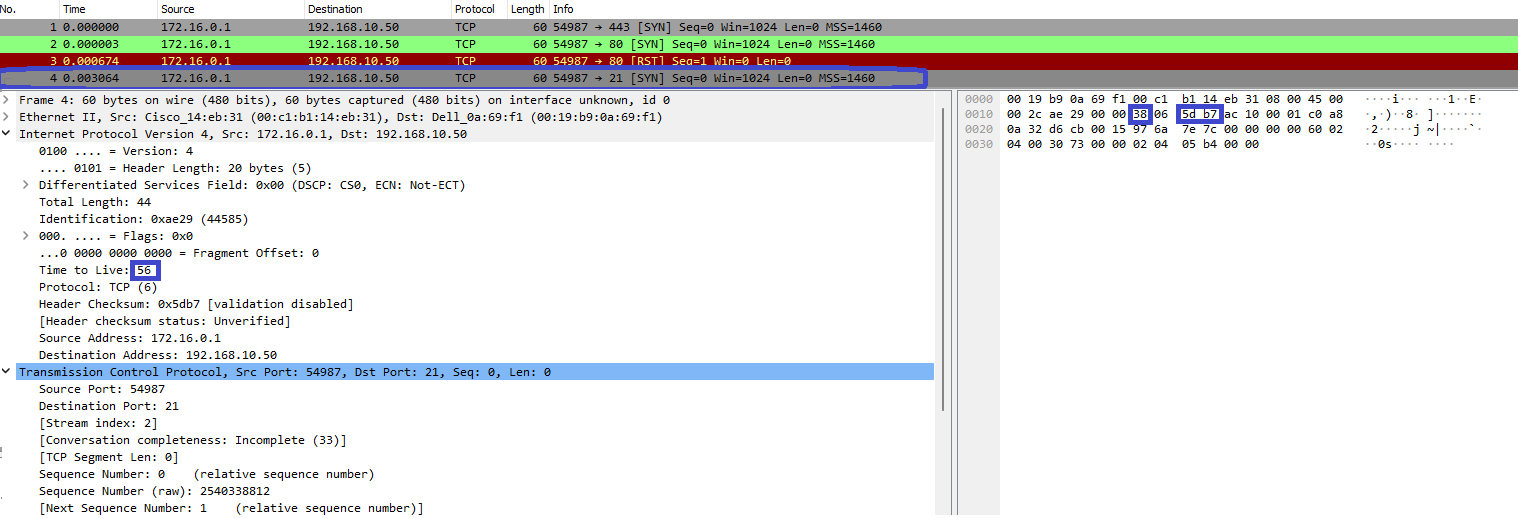}
\caption{Original Port Scan packet in Wireshark}
\label{origps}
\end{figure*}

\begin{figure*}
\centering
\includegraphics[width=0.95\textwidth]{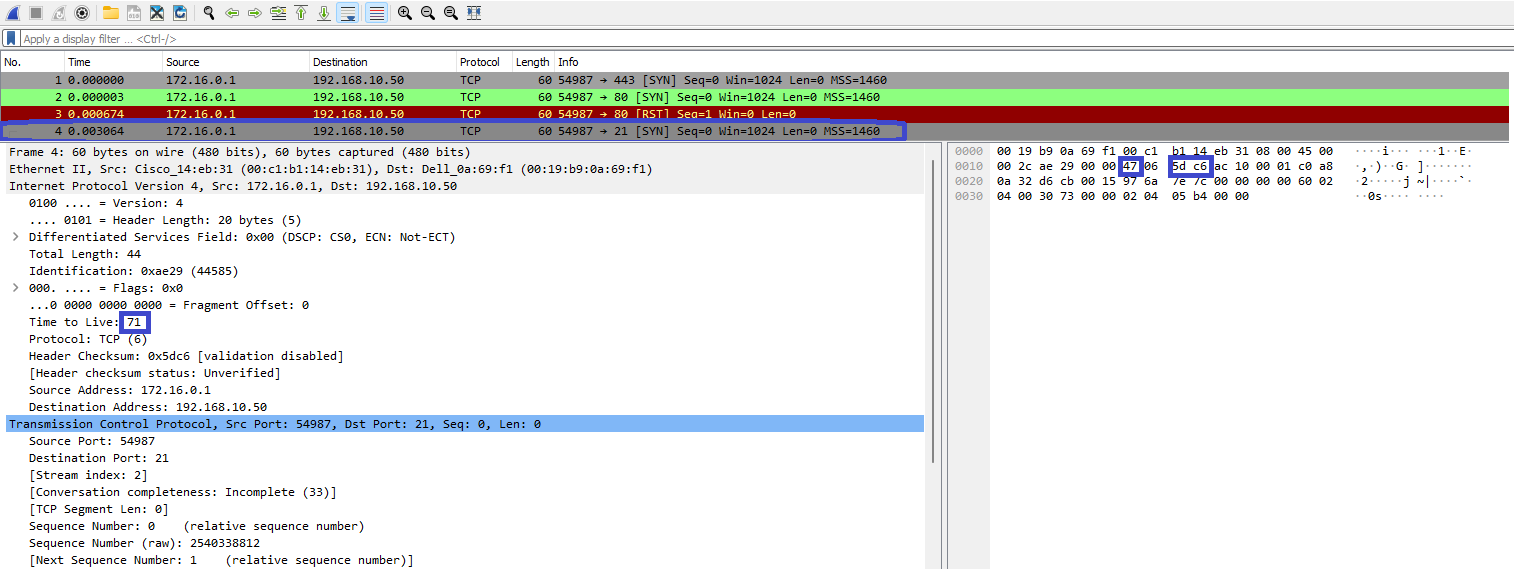}
\caption{Adversarial Port Scan packet in Wireshark}
\label{pertps}
\end{figure*}

We show two examples of successful perturbation by the \emph{Port Scan} agent. We plot the first 100 features of an original Port Scan packet and its respective adversarial sample. Figure~\ref{origpert} (a) shows an adversarial sample with a successful perturbation on the ninth feature (IP header byte number 9) and Figure~\ref{origpert} (b) demonstrates the successful addition of payload information in bytes 60-100. Both these actions, increasing the TTL values (directly affecting IP header byte numbers 9, 11, and 12) and adding payload (segment information), were amongst the top actions performed by the \emph{Port Scan} agent during testing. We used the SHAP values of the classifiers to determine any correlation between the agent's actions and the important feature(s) governing their decision boundaries. Figure~\ref{SHAP} (a--b) shows the SHAP values for the features in the RF and MLP classifiers, respectively, used to detect Port Scan attacks. Notably, IP header byte number 9 has the most significant contribution towards deciding the decision boundary for both models, which was also learned by the \emph{Port Scan} agent. Figures~\ref{origps} and \ref{pertps} show the Port Scan packets before and after successful perturbation of the TTL value, respectively, using the Wireshark packet analyzer.

\begin{figure*}
\centering
\includegraphics[width=0.95\textwidth]{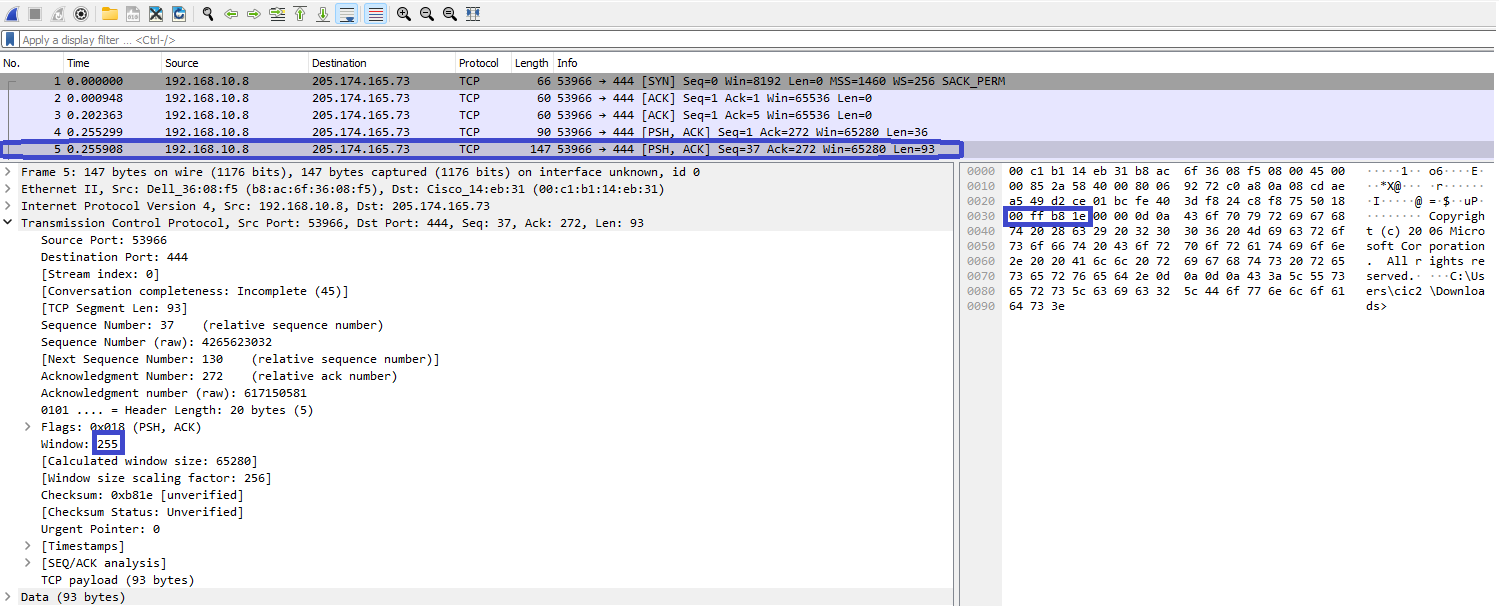}
\caption{Original Infiltration packet in Wireshark}
\label{originf}
\end{figure*}

\begin{figure*}
\centering
\includegraphics[width=0.95\textwidth]{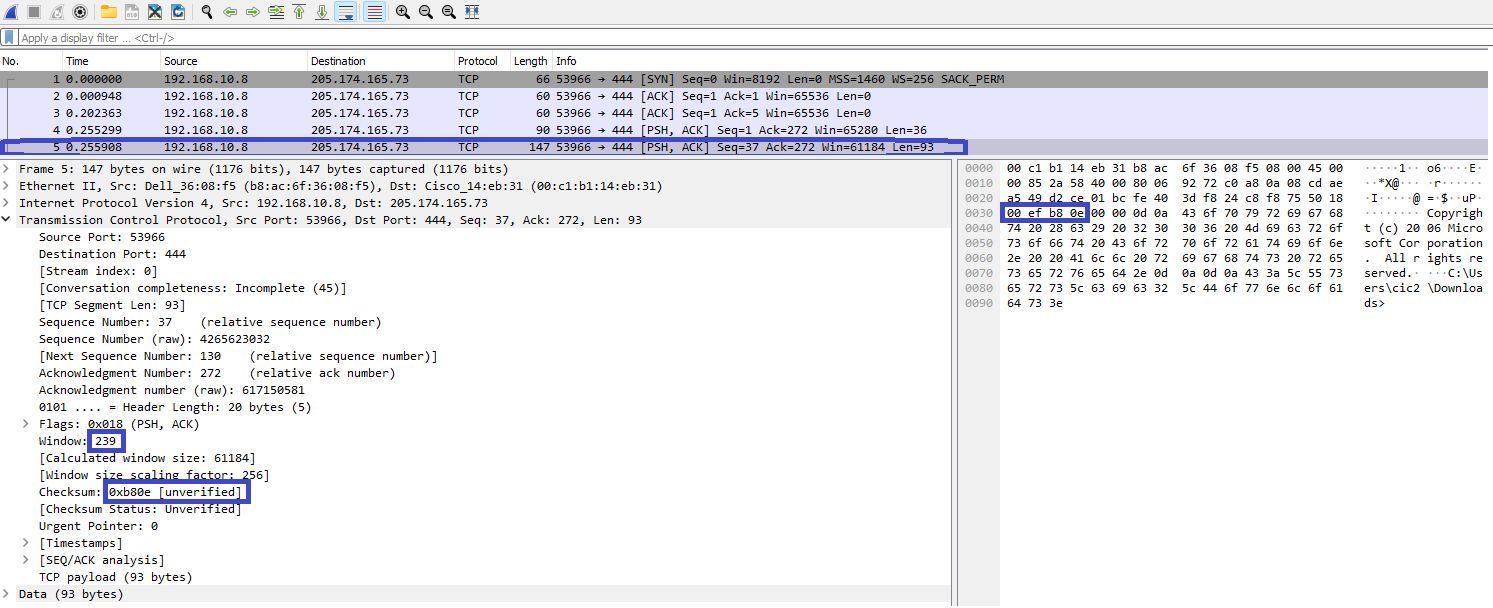}
\caption{Adversarial Infiltration packet in Wireshark}
\label{pertinf}
\end{figure*}

The DRL agents' performance analysis reveals that some attack types were less successful than others, as shown in Table~\ref{tab:ASR-CICIDS-2017}. The Infiltration attack type had the lowest success rate, likely due to its dissimilar feature values compared to the benign packets (see Figure~\ref{benignattack} (d)). Note that the Infiltration packets have unique feature values in bytes 61-1525, representing segment information, making it challenging to perturb these packets by modifying these features. Figure~\ref{origpert} (c) shows an example of a successful perturbation by the \emph{Infiltration} agent against an RF classifier. As seen in this figure, the agent was successful by making a modification in the window size feature value (directly impacting TCP header byte numbers 15 and 16). Figure~\ref{SHAP} (c) shows that the window size feature is the top contributor to the decision boundary for the RF classifier, which explains why perturbing window size value is amongst the top action choices for the \emph{Infiltration} agent. Figures~\ref{originf} and \ref{pertinf} show the original Infiltration attack packet and the DRL agent-generated sample with perturbed window size value, respectively, using Wireshark. These examples, amongst others, demonstrate that the DRL agents learned which key feature(s) to perturb for evading the classification boundary.

Notably, the early part of the TCP segment plays an important role in determining the MLP model's classification boundary (see Figure~\ref{SHAP} (d)). However, the early TCP segment information is not mutable for this attack type as that would compromise the functionality of the packet. Hence, the agent relied on adding payload (dead bytes) to these packets as an action choice and succeeded in some cases. An example of such a perturbation against an MLP classifier is shown in Figure~\ref{origpert} (d). The difference in the important features between classifiers also explains the \emph{Infiltration} agent's relatively higher success rate against the RF classifier compared to the MLP classifier. A similar performance analysis was conducted for the other DRL agents across all the classifiers. We found a similar correlation between the agent’s policy, the important features of the classifier, and the respective ASR.

\subsection{Statistical Analysis of the Successful Adversarial Packets}
Finally, we statistically analyze the successfully perturbed adversarial samples to determine their probabilistic difference from the original samples. We conduct the Kolmogorov-Smirnov test (K-S test) to quantify this difference. The two-sample K-S test compares the empirical cumulative distribution functions (eCDFs) of the perturbed adversarial sample and the original malicious sample to determine whether the former comes from the same distribution as the latter. It gives the maximum distance between the two eCDFs, also known as the K-S statistic (D)~\cite{gretton2012kernel}. If the calculated value of D is greater than the critical value for the specified significance level, then the null hypothesis is rejected, indicating that the samples do not come from the same distribution. We report the percentage of successful adversarial samples that were found to be out-of-distribution (OOD) in Tables~\ref{tab:OOD-CICIDS-2017} and \ref{tab:OOD-CICIDS-2018} at 95\% significance level. The former table shows the percentage of successfully perturbed adversarial samples from a different distribution than their original malicious samples in the CICIDS-2017 data set. The latter shows the percentage using the CICIDS-2018 data set.

We noted that the most successful adversary in both testing environments, the \emph{DDoS} agent, also generated the highest percentage of OOD samples. It can be inferred from Tables~\ref{tab:OOD-CICIDS-2017} and \ref{tab:OOD-CICIDS-2018} that the other agents did not need to make significant changes to the packets to fool the DT classifier, and as a result, they generated fewer OOD samples. We also noticed that a lower ASR value in Tables~\ref{tab:ASR-CICIDS-2017} and \ref{tab:ASR-CICIDS-2018} was correlated with a higher percentage of OOD samples, suggesting that these samples required more significant perturbations to evade the respective classifier. For example, the \emph{DoS} agent had to mainly rely on generating OOD samples to be successful against the SVM models. On average, over 45\% of the successful adversarial samples generated by the DRL agents across all testing models in both environments were OOD. In summary, our DRL-enabled adversarial network packet generation methodology, Deep PackGen, has shown encouraging results for generating OOD samples, which can inspire further research in this area to strengthen defenses against adversarial evolution.

\begin{table}[]
\centering
\caption{\% of out-of-distribution adversarial samples generated using CICIDS-2017 data}
\label{tab:OOD-CICIDS-2017}
\begin{tabular}{|c|ccccc|}
\hline
\multirow{3}{*}{Attack Type}    & \multicolumn{5}{c|}{Testing 1 Models}                                                                                                                                                                         \\ \cline{2-6} 
                                & \multicolumn{1}{c|}{\multirow{2}{*}{DT}} & \multicolumn{1}{c|}{\multirow{2}{*}{RF}} & \multicolumn{1}{c|}{\multirow{2}{*}{MLP}} & \multicolumn{1}{c|}{\multirow{2}{*}{DNN}} & \multirow{2}{*}{SVM}       \\
                                & \multicolumn{1}{c|}{}                    & \multicolumn{1}{c|}{}                    & \multicolumn{1}{c|}{}                     & \multicolumn{1}{c|}{}                     &                            \\ \hline
DoS                             & \multicolumn{1}{c|}{0}                   & \multicolumn{1}{c|}{25.21}               & \multicolumn{1}{c|}{50.61}                & \multicolumn{1}{c|}{48.15}                & 51.39                      \\ \hline
DDoS                            & \multicolumn{1}{c|}{85.86}               & \multicolumn{1}{c|}{82.61}               & \multicolumn{1}{c|}{85.11}                & \multicolumn{1}{c|}{74.19}                & 76.48                      \\ \hline
Web Attack                      & \multicolumn{1}{c|}{0.95}                & \multicolumn{1}{c|}{86.23}               & \multicolumn{1}{c|}{40.71}                & \multicolumn{1}{c|}{26}                   & 92.45                      \\ \hline
\multicolumn{1}{|l|}{Port Scan} & \multicolumn{1}{l|}{0.07}                & \multicolumn{1}{l|}{0.35}                & \multicolumn{1}{l|}{49.64}                & \multicolumn{1}{l|}{19.93}                & \multicolumn{1}{l|}{17.10} \\ \hline
Infiltration                    & \multicolumn{1}{c|}{0}                   & \multicolumn{1}{c|}{0.30}                & \multicolumn{1}{c|}{4.20}                 & \multicolumn{1}{c|}{37.16}                & 5.03                       \\ \hline
\end{tabular}
\end{table}

\begin{table}[]
\centering
\caption{\% of out-of-distribution adversarial samples generated using CICIDS-2018 data}
\label{tab:OOD-CICIDS-2018}
\begin{tabular}{|c|ccccc|}
\hline
\multirow{3}{*}{Attack Type} & \multicolumn{5}{c|}{Testing 2 Models}                                                                                                                                                                   \\ \cline{2-6} 
                             & \multicolumn{1}{c|}{\multirow{2}{*}{DT}} & \multicolumn{1}{c|}{\multirow{2}{*}{RF}} & \multicolumn{1}{c|}{\multirow{2}{*}{MLP}} & \multicolumn{1}{c|}{\multirow{2}{*}{DNN}} & \multirow{2}{*}{SVM} \\
                             & \multicolumn{1}{c|}{}                    & \multicolumn{1}{c|}{}                    & \multicolumn{1}{c|}{}                     & \multicolumn{1}{c|}{}                     &                      \\ \hline
DoS                          & \multicolumn{1}{c|}{0}                   & \multicolumn{1}{c|}{57.58}               & \multicolumn{1}{c|}{39.62}                & \multicolumn{1}{c|}{51.73}                & 76.06                \\ \hline
DDoS                         & \multicolumn{1}{c|}{61.14}               & \multicolumn{1}{c|}{71.48}               & \multicolumn{1}{c|}{52.55}                & \multicolumn{1}{c|}{77.76}                & 45.43                \\ \hline
Web Attack                   & \multicolumn{1}{c|}{23.43}               & \multicolumn{1}{c|}{81.68}               & \multicolumn{1}{c|}{75.92}                & \multicolumn{1}{c|}{76.06}                   & 36                \\ \hline
Infiltration                 & \multicolumn{1}{c|}{26.49}               & \multicolumn{1}{c|}{50.13}               & \multicolumn{1}{c|}{48.68}                & \multicolumn{1}{c|}{63}                   & 52.87                \\ \hline
\end{tabular}
\end{table}

\section{CONCLUSIONS AND FUTURE DIRECTIONS}
In this paper we presented the development of a generalized methodology for creating adversarial network packets that can evade ML-based NIDS. The methodology is aimed at finding (near-)optimal perturbations that can be made to malicious network packets while evading detection and retaining functionality for communication. We posed this constrained packet perturbation problem as a sequential decision-making problem and solved it using a DRL approach. The DRL-enabled solution framework, Deep PackGen, consists of three main components, namely packet-based data set creation, ML-based packet classification model development, and DRL-based adversarial sample generation. Raw packet capture files from publicly available data were used to conduct the experiments. The framework generated curated data sets containing forward network packets, which were used to train and test five different types of DRL agents. Each agent was tailored to a specific attack type and evaluated on various classifiers. Results show that the Deep PackGen framework is successful in producing adversarial packets with an average ASR of 66.4\% across all the classifiers in the network environment in which they were trained. The experimental results also show that the trained DRL agents produce an average ASR of 39.8\% across various tree-based and nonlinear models in a different network environment.

Below, we present a summary of the insights obtained from this study that can guide future investigations.
\begin{enumerate}
\item The DRL agents have a higher success rate in evading tree-based packet classification models like DT and RF compared to nonlinear classifiers such as SVM and DNN.
\item The success rate of the DRL agents in generating adversarial samples is directly related to the key features that govern the decision boundary of the classifier and whether these features could be changed without disrupting the packet's communication function.
\item Attacks that have feature values similar to those of benign traffic, such as DDoS and Port Scan, are more vulnerable to successful perturbation by an adversarial agent.
\item The policies learned by the DRL-agents are transferable to new network environments. 
\item The more complex the decision boundary of the classifier, the larger the magnitude of the perturbation required for evasion, resulting in out-of-distribution samples. 
\end{enumerate}

As regards future work, the Deep PackGen framework could be expanded to include more attack types and can be evaluated against different types of NIDS. Our methodology has shown encouraging results for generating OOD samples, which can be further investigated by the cybersecurity research community to model adversarial evolution and strengthen defenses against new types of attacks.

\section*{Acknowledgments}
This work was supported in part by the U.S. Military Academy (USMA) under Cooperative Agreement No. W911NF-22-2-0045, as well as the U.S. Army Combat Capabilities Development Command C5ISR Center under Support Agreement No. USMA21056. The views and conclusions expressed in this paper are those of the authors and do not reflect the official policy or position of the U.S. Military Academy, U.S. Army, U.S. Department of Defense, or U.S. Government.

\bibliographystyle{unsrtnat}
\bibliography{references}  






\end{document}